\documentclass[aip,graphicx,reprint,nofootinbib]{revtex4-1}

\usepackage{psfrag,graphicx}
\usepackage{dcolumn}
\usepackage{bm}
\usepackage{amsfonts,amssymb,amsmath}
\usepackage{bbold}
\usepackage{epstopdf}
\usepackage{xcolor}
\usepackage{graphicx}
\usepackage{color}

\begin{document}

\title{Majorana bound states in semiconducting nanostructures}



\author{Katharina Laubscher}
%\email{katharina.laubscher@unibas.ch}
\author{Jelena Klinovaja}

\affiliation{Department of Physics, University of Basel, Klingelbergstrasse 82, CH-4056 Basel, Switzerland}


\date{\today}

\begin{abstract}
In this Tutorial, we give a pedagogical introduction to Majorana bound states (MBSs) arising in semiconducting nanostructures. We start by briefly reviewing the well-known Kitaev chain toy model in order to introduce some of the basic properties of MBSs before proceeding to describe more experimentally relevant platforms. Here, our focus lies on simple `minimal' models where the Majorana wave functions can be obtained explicitly by standard methods. In a first part, we review the paradigmatic model of a Rashba nanowire with strong spin-orbit interaction (SOI) placed in a magnetic field and proximitized by a conventional $s$-wave superconductor. We identify the topological phase transition separating the trivial phase from the topological phase and demonstrate how the explicit Majorana wave functions can be obtained in the limit of strong SOI. In a second part, we discuss MBSs engineered from proximitized edge states of two-dimensional (2D) topological insulators. We introduce the Jackiw-Rebbi mechanism leading to the emergence of bound states at mass domain walls and show how this mechanism can be exploited to construct MBSs. Due to their recent interest, we also include a discussion of Majorana corner states in 2D second-order topological superconductors. This Tutorial is mainly aimed at graduate students---both theorists and experimentalists---seeking to familiarize themselves with some of the basic concepts in the field.
\end{abstract}

\maketitle %\maketitle must follow title, authors, abstract and \pacs

\section{Introduction}
\label{sec:intro}

In 1937, the Italian physicist Ettore Majorana proposed the existence of an exotic type of fermion---later termed a Majorana fermion---which is its own antiparticle.~\cite{Majorana1937} While the original idea of a Majorana fermion was brought forward in the context of high-energy physics,\cite{Elliott2015} it later turned out that emergent excitations with related properties can also be constructed in condensed matter systems.~\cite{Beenakker2013} Of particular interest in this context are so-called \emph{Majorana bound states} (MBSs) emerging at point-like defects in a special class of superconducting systems referred to as \emph{topological superconductors} (TSCs).\cite{Sato2016,Sato2017,Hasan2010} These MBSs are characterized by several intriguing properties: Firstly, similarly to their high-energy cousins, MBSs can be interpreted as being their own antiparticles in the sense that, in second-quantized language, the creation and annihilation operator associated with an MBS are equal to each other. This also immediately implies that an MBS carries both zero spin and zero charge. Secondly, MBSs appear at exactly zero energy and are separated from other, conventional quasiparticle excitations by a finite energy gap. For this reason, MBSs are also often referred to as Majorana zero modes (MZMs) and we will use the two terms interchangeably in the following. Thirdly---and probably most importantly---it was shown almost two decades ago that MBSs in a two-dimensional (2D) host material obey quantum exchange statistics that are neither fermionic nor bosonic.~\cite{Moore1991,Volovik1999,Read2000,Senthil2000,Ivanov2001,Volovik2003,Volovik2009} Rather, MBSs are an example of so-called \emph{non-Abelian anyons}. Loosely speaking, this means that exchanging two MBSs realizes a non-trivial rotation of the many-body ground state within a degenerate ground state subspace, with subsequent such rotations not necessarily commuting. This property makes non-Abelian anyons such as MBSs promising potential building blocks for topological quantum computers, where logical gates would then be performed by exchanging (`braiding') anyons.~\cite{Kitaev2003,Nayak2008}

\begin{figure}[t]
	\centering
	\includegraphics[width=\columnwidth]{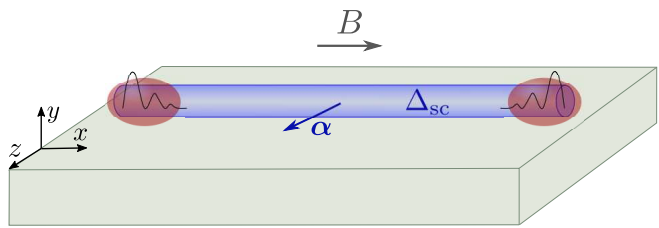}
	\caption{A nanowire (blue) placed on a superconducting substrate (green). The SOI vector $\boldsymbol\alpha$ is taken to lie along the $z$ axis and determines the spin quantization axis. A magnetic field $B$ is applied along the $x$ axis such that it is oriented perpendicularly to the SOI vector. By proximity, the substrate induces a superconducting pairing of strength $\Delta_{sc}$ in the nanowire. In the topological phase [see Eq.~(\ref{eq:top_criterion})], MBSs (red) emerge at the wire ends.
	}
	\label{fig:model_rashba}
\end{figure}

MBSs in condensed matter systems were first predicted to occur in the $\nu=5/2$ fractional quantum Hall state\cite{Moore1991} and in superconductors with an exotic spin-triplet pairing symmetry.~\cite{Volovik1999,Read2000,Senthil2000,Ivanov2001,Volovik2003,Volovik2009,Kitaev2001} Unfortunately, though, materials with intrinsic spin-triplet superconductivity turn out to be extremely rare. Crucially, it was later realized that an \emph{effective} spin-triplet pairing component can be engineered from a conventional spin-singlet $s$-wave superconductor when combined with a material exhibiting \emph{helical} bands, i.e., bands with spin-momentum locking.~\cite{Fu2008} With the realization of MBSs now suddenly seeming well within experimental reach, the field has witnessed a veritable explosion. Following the earliest proposals based on proximitized topological insulator (TI) edge or surface states,~\cite{Fu2008,Fu2009} MBSs have subsequently been proposed to emerge in 2D semiconducting quantum wells~\cite{Alicea2010,Sau2010a,Sato2009,Sato2010} and one-dimensional (1D) semiconductor nanowires with strong Rashba spin-orbit interaction (SOI),\cite{Lutchyn2010,Sau2010b,Oreg2010,Stanescu2011} chains of magnetic adatoms deposited on a superconductor,\cite{Choy2011,Klinovaja2013c,Vazifeh2013,Braunecker2013,Nadj-Perge2013,Pientka2013} TI nanowires,~\cite{Cook2011,Cook2012,Legg2021} graphene-based structures such as carbon nanotubes and nanoribbons,~\cite{Klinovaja2012c,Egger2012,Klinovaja2013a,Sau2013,Dutreix2014,Marganska2018,Desjardins2019} planar Josephson junctions,\cite{Pientka2017,Hell2017,Fornieri2019,Ren2019,Dartiailh2021} and many more.

Due to its expected experimental feasibility, the Rashba nanowire model~\cite{Oreg2010,Lutchyn2010,Sau2010b,Stanescu2011} is one of the most well-explored proposals among the above. Here, a semiconducting nanowire with strong Rashba SOI is placed in a magnetic field and proximitized by a conventional $s$-wave superconductor, see Fig.~\ref{fig:model_rashba}. The magnetic field, in combination with the strong SOI, leads to the emergence of a helical regime with two counterpropagating bulk modes carrying opposite spin projections. Since all of the required ingredients are in principle readily available in the laboratory, this proposal has triggered significant experimental activity, culminating in a series of works measuring zero-bias conductance peaks consistent with the signatures expected from MBSs.~\cite{Das2012,Rokhinson2012,Deng2012,Williams2012,Lee2012,Deng2018,DeMoor2018,Mourik2012,Churchill2013,Deng2016,Vaitiekenas2018} However, it was soon realized that very similar zero-bias peaks can also appear due to alternative, non-topological mechanisms in the absence of MBSs.\cite{Prada2020,abs1,abs2,abs3,abs4,abs5,liu2017andreev,reeg2018zero,liu2019conductance,alspaugh2020volkov,abs6,abs7,abs8,abs10,Dmytruk2020,Yu2020,Kayyalha2020,Valentini2020} 
As such, irrefutable experimental proof of the presence of MBSs has not been obtained up to date.

In spite of---or maybe exactly because of---these unresolved issues, the field of MBSs is a highly active and rapidly evolving research area. The goal of this Tutorial is to familiarize graduate students---both theorists and experimentalists---with some of the most well-known proposed realizations of MBSs in semiconducting nanostructures. We assume the reader to be familiar with elementary quantum mechanics and the formalism of second quantization. Furthermore, some basic knowledge of the Bogoliubov-de Gennes (BdG) formalism used to treat systems with superconducting order at the mean-field level is highly beneficial. Readers unfamiliar with this concept may for example consult Ref.~\onlinecite{Bernevig2013} for a pedagogical introduction. Throughout this Tutorial, we try to expose the relevant physical properties of the considered systems using only the simplest mathematical tools. In particular, we will not introduce the concept of topological invariants and the elaborate mathematical framework underlying the general theory of topological phases of matter. For this and other relevant aspects that are not covered in this Tutorial, we refer the reader to several excellent comprehensive reviews of the field.~\cite{Alicea2012,Beenakker2013,Aguado2017,Lutchyn2018,Stanescu2013,DasSarma2015,Sato2017,Sato2016,Leijnse2012,Elliott2015,Pawlak2019,Leon2021,Jack2021,vonOppen2017}

The Tutorial is organized as follows: In Sec.~\ref{sec:prelim}, we introduce some basic notions related to MBSs that will reappear frequently throughout the entire Tutorial. Using the Kitaev chain toy model,~\cite{Kitaev2001} we demonstrate how isolated MBSs can emerge in a superconducting system and how their pinning to zero energy is guaranteed by particle-hole symmetry. We explain how an MBS can intuitively be pictured as `half' a fermionic zero mode and how two spatially separated MBSs encode a non-local fermionic degree of freedom, leading to a two-fold ground state degeneracy of the system. After this introductory part focusing on a toy model, we turn our attention to more physical systems. First, in Sec.~\ref{sec:nanowires}, we discuss the paradigmatic model of a Rashba nanowire proximitized by a conventional $s$-wave superconductor in the presence of a magnetic field. We identify the topological phase transition separating the trivial phase from the topological phase with a pair of MBSs at the wire ends, demonstrating how the exact MBS wave functions can be obtained in the limit of strong Rashba SOI. Furthermore, we also introduce the concept of synthetic SOI. In Sec.~\ref{sec:TI}, we then turn our attention to proximitized TI edge states as an alternative platform for MBSs. We introduce the Jackiw-Rebbi mechanism~\cite{Jackiw1976,Jackiw1981} leading to the emergence of bound states at mass domain walls and show how this mechanism can be exploited to construct MBSs. For completeness, we comment on a selection of other (quasi-)1D platforms hosting MBSs in Sec.~\ref{sec:others}. Finally, due to their recent interest, Sec.~\ref{sec:MCSs} is dedicated to 2D higher-order topological superconductors hosting so-called Majorana corner states. We conclude in Sec.~\ref{sec:conclusions}.

\section{Preliminaries}
\label{sec:prelim}

\begin{figure*}[tb]
	\centering
	\includegraphics[width=0.75\textwidth]{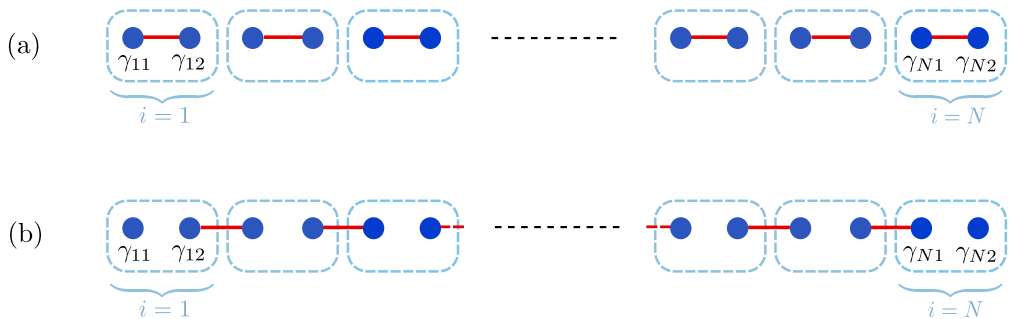}
	\caption{Pictorial representation of the Kitaev chain model [see Eq.~(\ref{eq:Kitaev_chain})] in two limiting cases. For illustrative purposes, the physical fermion living on site $i$ (light blue box) is split up into two Majorana operators $\gamma_{i1}$, $\gamma_{i2}$ (blue dots) according to Eq.~(\ref{eq:Majorana_operators}). (a) $\Delta=t=0$. In this case, the chain is in the trivial phase, where Majorana operators belonging to the same site are coupled (red line). (b) $\Delta=t>0$ and $\mu=0$. In this case, the chain is in the topologically non-trivial phase, where Majorana operators belonging to neighboring sites are coupled (red line). This results in two Majorana operators $\gamma_{11}$ and $\gamma_{N2}$ at the ends of the chain not entering the Hamiltonian.}
	\label{fig:kitaev_chain}
\end{figure*}

In this section, we introduce some basic notions needed to understand the emergence of Majorana zero modes in condensed matter systems. The simple ideas developed here will frequently reappear throughout the entire Tutorial.

Let us start by recalling that, within the formalism of second quantization, a single species of electrons is described by creation and annihilation operators $c^\dagger$, $c$ satisfying the canonical anticommutation relations $\{c^\dagger,c^\dagger\}=\{c,c\}=0$, $\{c,c^\dagger\}=1$. If we now want to construct a quasiparticle excitation that behaves as its own antiparticle, $\gamma=\gamma^\dagger$, it is clear that it has to take the form of an equal-weight linear combination of electronic creation and annihilation operators, $\gamma=e^{i\phi}c+e^{-i\phi}c^\dagger$ for some phase $\phi$. We can thus already guess that the most natural place to look for MBSs is in superconductors. Indeed, the quasiparticle excitations of a superconductor (also referred to as Bogoliubov quasiparticles) are linear combinations of electronic creation and annihilation operators.~\cite{Bardeen1957} However, it is not enough to just take a plain $s$-wave superconductor: In such a conventional superconductor, the Bogoliubov quasiparticles take the form $\gamma=uc_\sigma+vc_{\bar\sigma}^\dagger$ for some complex coefficients $u$ and $v$ and with spin $\sigma\in\{\uparrow,\downarrow\}$,\footnote{Throughout this Tutorial, we use the shorthand notation $\bar\uparrow=\downarrow$, $\bar\downarrow=\uparrow$. A similar notation will be used for other quantum numbers as well.} and where we have omitted any other degrees of freedom that might be present. Clearly, such a particle can never be its own antiparticle due to the mismatching spin indices. Fortunately, there are ways to circumvent this problem: One can, for example, consider more exotic types of superconductors beyond the standard Bardeen-Cooper-Schrieffer (BCS) theory.\cite{Bardeen1957}  
Indeed, it turns out that \emph{spinless $p$-wave} superconductors are the simplest platforms capable of hosting MBSs. Even though this type of superconductivity is unlikely to occur naturally, we will see later that an effective spinless $p$-wave component can also be realized in heterostructures combining conventional $s$-wave superconductors with suitable other materials, such as, e.g., semiconductors with strong SOI. As such, studying of the simplest spinless $p$-wave superconductor can give important conceptual insights into the emergence of MBSs. We will therefore devote the remainder of this introductory section to a toy model based on spinless fermions with intrinsic $p$-wave pairing before moving to more physical models in the next sections.

Explicitly, we consider a 1D chain of spinless fermions described by the Hamiltonian\cite{Lieb1961}
\begin{equation}
	H=\sum_{i=1}^{N-1}(-tc_i^\dagger c_{i+1}+\Delta c_ic_{i+1}+\mathrm{H.c.})-\mu\sum_{i=1}^{N}c_i^\dagger c_i.
	\label{eq:Kitaev_chain}
\end{equation}
Here, $N$ is the number of sites, $t\geq0$ is the nearest-neighbor hopping amplitude, $\Delta\geq0$ is the superconducting pairing potential, which we have taken to be real for simplicity, and $\mu$ is the chemical potential. This Hamiltonian was popularized by Ref.~\onlinecite{Kitaev2001} and is therefore often referred to as the \emph{Kitaev chain}. Following the original work, we find it convenient to introduce new operators $\gamma_{i1}$, $\gamma_{i2}$ for $i\in\{1,...,N\}$ defined via
\begin{equation}
	c_i=\frac{1}{2}(\gamma_{i1}+i\gamma_{i2}),\qquad c_i^\dagger=\frac{1}{2}(\gamma_{i1}-i\gamma_{i2}).
	\label{eq:Majorana_operators}
\end{equation}
Loosely speaking, this can be thought of as separating the electron operator into its real and imaginary parts. The inverse relation of the above transformation reads
\begin{equation}
	\gamma_{i1}=c_i^\dagger+c_i,\qquad\gamma_{i2}=i(c_i^\dagger-c_i),
\end{equation}
and it can be checked that the new operators $\gamma_{i1}$, $\gamma_{i2}$ satisfy the relations
\begin{equation}
	\{\gamma_{i\alpha},\gamma_{j\beta}\}=2\delta_{ij}\delta_{\alpha\beta},\qquad\gamma_{i\alpha}^\dagger=\gamma_{i\alpha}.
\end{equation}
From the second relation, we thus see that the $\gamma_{i\alpha}$ correspond to Majorana operators in the sense discussed above. 

Rewritten in terms of the Majorana operators, the Hamiltonian given in Eq.~(\ref{eq:Kitaev_chain}) takes the form
\begin{align}
H&=\frac{i}{2}\sum_{i=1}^{N-1}\left[(\Delta+t)\gamma_{i2}\gamma_{(i+1)1}+(\Delta-t)\gamma_{i1}\gamma_{(i+1)2}\right]	\nonumber\\&\quad\,-\frac{\mu}{2}\sum_{i=1}^N(i\gamma_{i1}\gamma_{i2}+1).
\end{align}
While this Hamiltonian is rather complicated in its full form, a lot of insight can be gained by focusing on certain limiting cases. For future reference, let us start by briefly looking at the trivial case $\Delta=t=0$, $\mu<0$, where the chain simply consists of uncoupled sites. The Hamiltonian then takes the form
\begin{equation}
H=-\mu\sum_{i=1}^{N}c_i^\dagger c_i=-\frac{\mu}{2}\sum_{i=1}^N(i\gamma_{i1}\gamma_{i2}+1).
\end{equation}
We now see that this corresponds to the case where the two Majorana operators corresponding to a physical fermion are paired up, see Fig.~\ref{fig:kitaev_chain}(a). The system is fully gapped as adding a fermion costs a finite energy $-\mu$. This phase is called the (topologically) trivial phase and does not host any MBSs.

A more interesting situation arises for $\Delta=t$ and $\mu=0$. In this case, the Hamiltonian reduces to 
\begin{equation}
	H=it\sum_{i=1}^{N-1}\gamma_{i2}\gamma_{(i+1)1}.
	\label{eq:Kitaev_chain_nontrivial}
\end{equation}
Pictorially, this now corresponds to the case where Majorana operators belonging to neighboring sites are paired up, see Fig.~\ref{fig:kitaev_chain}(b). It is therefore insightful to rewrite the Hamiltonian given in Eq.~(\ref{eq:Kitaev_chain_nontrivial}) in terms of new fermionic operators
\begin{equation}
d_i=\frac{1}{2}(\gamma_{i2}+i\gamma_{(i+1)1}),\quad d_i^\dagger=\frac{1}{2}(\gamma_{i2}-i\gamma_{(i+1)1}),
\end{equation} 
leading to
\begin{equation}
	H=2t\sum_{i=1}^{N-1} \left(d_i^\dagger d_i-\frac{1}{2}\right).
\end{equation}
As such, we see that the bulk of the system is still gapped, i.e., adding a fermion of type $d$ costs a finite energy $2t$. However, the two Majorana operators $\gamma_{11}$ at the left end of the chain and $\gamma_{N2}$ at the right end of the chain do not enter the Hamiltonian at all. This implies $[H,\gamma_{11}]=[H,\gamma_{N2}]=0$, i.e., the two outermost sites of the chain now host two isolated zero-energy modes satisfying the Majorana property $\gamma_{11}=\gamma_{11}^\dagger$, $\gamma_{N2}=\gamma_{N2}^\dagger$. These are exactly the MBSs we were looking for, and we say that the system is now in the topologically nontrivial phase (topological phase) with one MBS at each end of the chain. As becomes clear from the pictorial representation in Fig.~\ref{fig:kitaev_chain}(b), we can think of each of the two MBSs as `half' a physical fermion. This also suggests that the two MBSs can be combined to form a single fermionic zero mode, 
\begin{equation}
d_0=\frac{1}{2}(\gamma_{11}+i\gamma_{N2}),\quad d_0^\dagger=\frac{1}{2}(\gamma_{11}-i\gamma_{N2}).
\end{equation} 
Since the constituent MBSs are localized far away from each other at opposite ends of the chain, this fermionic zero mode is highly delocalized. Furthermore, this non-local fermionic state can be filled or emptied at zero energy cost, leading to the presence of two degenerate ground states differing in their fermion number. This two-fold ground state degeneracy, together with the non-Abelian exchange statistics of MBSs, lies at the heart of the idea that MBSs can be used for topological quantum computation. While a detailed discussion of how MBSs can be used to perform topologically protected quantum gates is beyond the scope of this Tutorial, we refer the interested reader to several excellent reviews covering this topic.~\cite{Alicea2012,Aguado2017,DasSarma2015,Beenakker2013,Leijnse2012,Elliott2015}

While the above discussion focused on two limiting cases corresponding to the fully dimerized situations shown in Fig.~\ref{fig:kitaev_chain}, the qualitative properties of the trivial and nontrivial phase persist also if one deviates from these fine-tuned points. Indeed, the number of MBSs at a given end of the chain is directly related to a \emph{topological invariant}, meaning that it is robust under continuous changes of the system parameters as long as the bulk gap remains open and none of the protecting symmetries are broken. More generally, the theory of topological phases of matter states that topologically inequivalent phases are separated by closing points of the bulk gap and differ by the values of their topological invariant. In the case of the Kitaev chain, this invariant takes the values 0 (trivial phase) or 1 (topological phase) and gives the number of MBSs at one end of the chain.\footnote{More explicitly, when taking particle-hole symmetry as the only symmetry of the model, the Kitaev chain belongs to the class D in the symmetry classification of topological phases of matter,\cite{Ryu2010,Chiu2016} which has a $\mathbb{Z}_2$ topological invariant in one dimension. Via the bulk-boundary correspondence, the value of this topological invariant is directly related to the number of MBSs at a given end of the chain.}

\begin{figure}[tb]
	\centering
	\includegraphics[width=0.8\columnwidth]{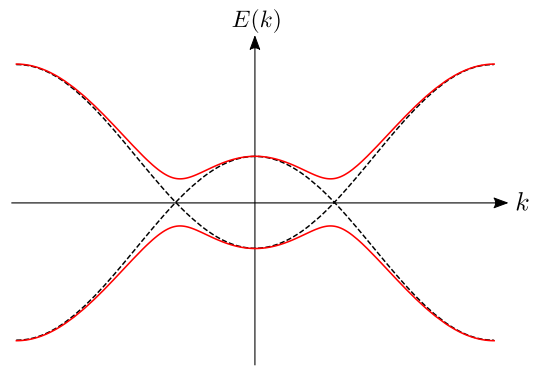}
	\caption{Bulk spectrum of the Kitaev chain, see Eq.~(\ref{eq:kitaev_bulk}). In the absence of superconductivity, i.e., for $\Delta=0$, the spectrum is gapless (black dashed lines). Any finite $\Delta$ leads to a fully gapped spectrum (red solid lines) for any $\mu$ within the normal-state band. In this example, we have used $\mu/t=-1$ and $\Delta=0$ for the black dashed line ($\Delta/t=0.3$ for the red solid line).
	}
	\label{fig:kitaev_dispersion}
\end{figure}

The above statement motivates us to look at the bulk spectrum of the Kitaev chain. Assuming periodic boundary conditions for the moment, we can rewrite the Hamiltonian given in Eq.~(\ref{eq:Kitaev_chain}) in momentum space. In order to account for the superconducting term, we resort to the standard BdG description employing the Nambu spinor $C_k^\dagger=(c_k^\dagger,c_{-k})$. The bulk Hamiltonian then takes the form $H=\frac{1}{2}\sum_kC_k^\dagger\mathcal{H}(k)C_k$ with
\begin{equation}
\mathcal{H}(k)=(-2t\cos{k}-\mu)\,\eta_z-2\Delta\sin{k}\,\eta_y.
\end{equation}
Here, $\eta_i$ for $i\in\{x,y,z\}$ are Pauli matrices acting in particle-hole space. It is important to note that, as for any mean-field BdG Hamiltonian in the Nambu representation, the electron and hole components of $\mathcal{H}(k)$ are not independent.\cite{Bernevig2013} More specifically, $\mathcal{H}(k)$ satisfies the so-called \emph{particle-hole symmetry}
\begin{equation}
U_C\mathcal{H}(k)U_C^{-1}=-\mathcal{H}^*(-k)
\label{eq:ph_symm}
\end{equation}
with $U_C=\eta_x$.\cite{Ryu2010} This symmetry guarantees that the spectrum of $\mathcal{H}(k)$ is symmetric around zero in the sense that for any eigenstate $\varphi_E(k)$ at energy $+E(k)$ there is also an eigenstate $U_C\varphi^*_E(-k)$ at energy $-E(-k)$. Explicitly, we find that the bulk energy spectrum is given by
\begin{equation}
E^2(k)=(2t\cos{k}+\mu)^2+4\Delta^2\sin^2{k}.
\label{eq:kitaev_bulk}
\end{equation}
We show an example plot of this energy spectrum in Fig.~\ref{fig:kitaev_dispersion}. It can easily be checked that, for any finite $\Delta$, the bulk spectrum is fully gapped except for $\mu=\pm 2t$. Thus, these points mark the possible transition between topologically distinct phases. Starting from the fully dimerized case shown in Fig.~\ref{fig:kitaev_chain}(b), we can then infer that the Kitaev chain remains in the topologically nontrivial phase for any finite $\Delta$ and any chemical potential $|\mu|<2t$, i.e., for any $\mu$ within the normal-state band. This also means that the two MBSs persist for any set of parameters within this range. However, the MBSs will generally cease to be perfectly localized to only the outermost sites of the chain. Instead, their spatial profile will decay exponentially into the bulk. This also means that in a chain of finite length, the two MBSs at opposite ends of the chain will acquire a finite overlap with each other, thereby lifting the exact two-fold ground state degeneracy discussed above. However, the energy splitting between the two ground states of opposite fermion parity is exponentially suppressed with the system length, meaning that our simplified picture introduced above remains valid in the limit of long chains.

An intuitive explanation for the stability of the MBSs can be given via a simple symmetry argument. This is easiest to understand when we first consider a semi-infinite system with a single end. In the topological phase, a single zero-energy MBS is located at this end [see, e.g., Fig.~\ref{fig:kitaev_chain}(b)]. Furthermore, the particle-hole symmetry discussed above can easily be recast to real space and also applies in the case of a finite system. This means that the spectrum of the BdG Hamiltonian is symmetric around zero energy, or, equivalently, the creation of a Bogoliubov quasiparticle at energy $E$ is equivalent to the annihilation of a quasiparticle (i.e., the creation of a quasihole) at energy $-E$. It is now straightforward to see that a single isolated MBS at $E=0$ cannot be removed from zero energy without violating particle-hole symmetry. This argument also remains valid for a long but finite chain: Since the overlap between the two MBSs is exponentially suppressed with the system size, it can become negligibly small for a sufficiently long chain. In this case, neither of the MBSs can be removed from zero energy by continuous changes of the system parameters or arbitrary particle-hole symmetric local perturbations as long as the bulk gap remains open. The exponential suppression of the energy splitting between the two degenerate ground states is also referred to as the topological protection of the ground state degeneracy.

Last but not least, let us stress once again that the Kitaev chain should merely be viewed as a toy model. Firstly, electrons in solids naturally come with a spin degree of freedom, and secondly, intrinsic spin-triplet superconductors are very rare to almost non-existent in nature. It is therefore natural to ask whether there exist ways to enable topological superconducting phases in heterostructures based on conventional $s$-wave superconductors. Fortunately, this is indeed possible when so-called \emph{helical} bands---i.e., bands with spin-momentum locking---are present. Indeed, when a conventional $s$-wave pairing term is projected onto a helical band, an effective spinless $p$-wave component emerges.\footnote{For an explicit demonstration of how this happens, see for example Ref.~\onlinecite{Alicea2012}.}

In the remainder of this Tutorial, we will explore some of the most well-known realizations of MBSs based on conventional $s$-wave superconductors. Our main focus lies on proximitized Rashba nanowires in the presence of a magnetic field (Sec.~\ref{sec:nanowires}) as well as proximitized topological insulator edge states (Sec.~\ref{sec:TI}). In light of our previous discussion of the Kitaev chain, our main strategy will be to first identify the topologically inequivalent phases of the models under consideration by looking for closing points of the bulk gap. Subsequently, we will derive the conditions under which MBSs emerge by explicitly solving for localized zero-energy wave function solutions of the corresponding BdG equations.

\section{Majorana bound states in nanowires}
\label{sec:nanowires}

In this section, we study the paradigmatic example of MBSs arising in semiconducting Rashba nanowires in the presence of a magnetic field and proximity-induced $s$-wave superconductivity.\cite{Lutchyn2010,Oreg2010,Sau2010b,Stanescu2011} Due to its expected experimental feasibility, this setup is often considered one of the most promising proposals for the realization of MBSs. Indeed, over the last decade, significant progress has been reported in experiments based on InAs or InSb nanowires.\cite{Mourik2012,Das2012,Rokhinson2012,Churchill2013,Deng2016} In the following, we present the theoretical arguments leading to the emergence of MBSs and demonstrate how the explicit MBS wave functions can be obtained in the limit of strong SOI. For this, we will closely follow the methods developed in Ref.~\onlinecite{Klinovaja2012}.

\begin{figure}[tb]
	\centering
	\includegraphics[width=0.7\columnwidth]{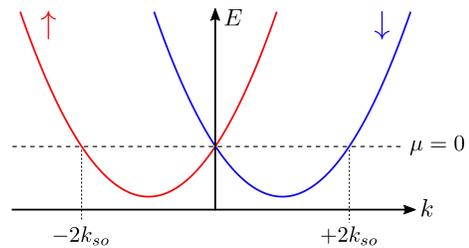}
	\caption{Bulk spectrum of a bare Rashba nanowire, see Eq.~(\ref{eq:spectrum_bare}). The spectrum consists of two parabolas---one for spin up and one for spin down---shifted by $\pm k_{so}$. At $\mu=0$, the Fermi momenta are given by $k=0$ and $k=\pm 2k_{so}$.}
	\label{fig:spectrumRashba}
\end{figure}

\subsection{MBSs in Rashba nanowires: Minimal model}

Our starting point is a 1D Rashba nanowire that is taken to be oriented along the $x$ axis, see Fig.~\ref{fig:model_rashba}. The bare nanowire is described by the Hamiltonian $H_0=H_{kin}+H_{SOI}$, where $H_{kin}$ denotes the kinetic contribution to the Hamiltonian and $H_{SOI}$ denotes the Rashba SOI. The SOI is characterized by the spin-orbit vector $\boldsymbol\alpha$, which is oriented perpendicular to the nanowire and defines our spin quantization axis. For concreteness, we assume that $\boldsymbol\alpha$ points along the $z$ axis, see again Fig.~\ref{fig:model_rashba}. Explicitly, the kinetic term takes the form
\begin{equation}
H_{kin}=\sum_{\sigma}\int dx\,\Psi_\sigma^\dagger(x)\left(-\frac{\hbar^2\partial_x^2}{2m}-\mu\right)\Psi_\sigma(x),
\label{eq:Hkin}
\end{equation}
where $\Psi_\sigma(x)$ [$\Psi_\sigma^\dagger(x)$] destroys (creates) an electron of spin $\sigma\in\{\uparrow,\downarrow\}$ at position $x$ in the nanowire, $m$ denotes the effective electron mass, and $\mu$ denotes the chemical potential. The SOI term is given by
\begin{equation}
H_{SOI}=-i\alpha\sum_{\sigma,\sigma'}\int dx\,\Psi_\sigma^\dagger(x)(\sigma_z)_{\sigma\sigma'}\partial_x\Psi_{\sigma'}(x),
\label{eq:Hsoi}
\end{equation}
where $\sigma_i$ for $i\in\{x,y,z\}$ are Pauli matrices acting in spin space and $\alpha>0$ denotes the strength of the SOI. Note that in this description, the chemical potential is measured relative to the SOI energy $E_{so}=m\alpha^2/(2\hbar^2)$. For a translationally invariant nanowire, $H_0$ can be written in momentum space as $H_0=\sum_k \Psi_k^\dagger\mathcal{H}_0(k)\Psi_k$, where we have defined $\Psi_k^\dagger=(\psi_{k\uparrow}^\dagger,\psi_{k\downarrow}^\dagger)$ and the Hamiltonian density $\mathcal{H}_0(k)$ is given by
\begin{equation}
	\mathcal{H}_0(k)=\frac{\hbar^2k^2}{2m}-\mu+\alpha k \sigma_z.
\end{equation}
The bulk spectrum of $H_0$ is readily found to be
\begin{equation}
E_{0,\pm}(k)=\frac{\hbar^2k^2}{2m}-\mu\pm\alpha k
\label{eq:spectrum_bare}
\end{equation}
and corresponds to two parabolas---one for each spin species---shifted relative to each other, see Fig.~\ref{fig:spectrumRashba}. In the special case $\mu=0$, the Fermi points are given by $k=0$ and $k=\pm2k_{so}$, where the spin-orbit momentum $k_{so}$ is given by $k_{so}=m\alpha/\hbar^2$. In general, for a small but possibly finite $\mu$, we will call the branches close to $k=0$ ($k=\pm 2k_{so}$) the interior (exterior) branches of the spectrum.

Let us now additionally consider the effect of a magnetic field of strength $B$ oriented perpendicular to the SOI vector. For concreteness, we take the magnetic field to be oriented along the $x$ axis in the following, see Fig.~\ref{fig:model_rashba}. The magnetic field gives rise to the Zeeman term
\begin{equation}
H_{Z}=\Delta_Z\sum_{\sigma,\sigma'}\int dx\,\Psi_\sigma^\dagger(x)(\sigma_x)_{\sigma\sigma'}\Psi_{\sigma'}(x)
\end{equation}
with $\Delta_Z=g\mu_BB/2$, where $g$ denotes the $g$-factor of the nanowire and $\mu_B$ the Bohr magneton. In the following, we will assume $\Delta_Z\geq 0$.

Finally, if the nanowire is brought in tunnel-contact with a bulk $s$-wave superconductor, we can account for the proximity-induced superconductivity via
\begin{equation}
H_{sc}=\frac{\Delta_{sc}}{2}\sum_{\sigma,\sigma'}\int dx\,\Psi_\sigma(x)(i\sigma_y)_{\sigma\sigma'}\Psi_{\sigma'}(x)+\mathrm{H.c.},
\end{equation}
where we take $\Delta_{sc}$ to be real and non-negative for simplicity. The total Hamiltonian that we will be concerned with in the remainder of this section is then given by $H=H_0+H_Z+H_{sc}$.

Assuming a translationally invariant system for the moment, we can rewrite $H$ in momentum space. It then takes the form $H=\frac{1}{2}\sum_k \Psi_k^\dagger\mathcal{H}(k)\Psi_k$, where we have introduced the Nambu spinor $\Psi_k^\dagger=(\psi_{k\uparrow}^\dagger,\psi_{k\downarrow}^\dagger,\psi_{-k\uparrow},\psi_{-k\downarrow})$ and the Hamiltonian density is given by
\begin{equation}
\mathcal{H}(k)=\left(\frac{\hbar^2k^2}{2m}-\mu\right)\eta_z+\alpha k \sigma_z+\Delta_Z\eta_z\sigma_x+\Delta_{sc}\eta_y\sigma_y.
\end{equation}
Here, $\eta_i$ for $i\in\{x,y,z\}$ are Pauli matrices acting in particle-hole space. Like any BdG Hamiltonian, this Hamiltonian is particle-hole symmetric as expressed by Eq.~(\ref{eq:ph_symm}). As discussed in Sec.~\ref{sec:prelim}, it is exactly this particle-hole symmetry that is responsible for pinning any isolated MBS---if present---to exactly zero energy. We will come back to this point when we explicitly solve for MBSs in Subsec.~\ref{subsec:MBSs_Rashba}.\footnote{Furthermore, if and only if $\Delta_Z=0$, the Hamiltonian also has time-reversal symmetry expressed by \begin{equation}U_T\mathcal{H}(k)U_T^{-1}=\mathcal{H}^*(-k)\label{eq:tr_symm}\end{equation} with $U_T=i\sigma_y$.\cite{Ryu2010,Chiu2016} However, as we will see later, the existence of MBSs in this minimal model necessarily requires that time-reversal symmetry is broken.}

Via a straightforward eigenvalue calculation, the spectrum of $\mathcal{H}(k)$ is found to be
\begin{align}
E_\pm^2(k)&=\left(\frac{\hbar^2 k^2}{2m}-\mu\right)^2+\alpha^2k^2+\Delta_Z^2+\Delta_{sc}^2\label{eq:spectrum}\\&\hspace{5mm}\pm 2\sqrt{\left(\frac{\hbar^2 k^2}{2m}-\mu\right)^2(\Delta_Z^2+\alpha^2k^2)+\Delta_Z^2\Delta_{sc}^2}.\nonumber
\end{align}
By direct inspection, one can show that, for any finite $\Delta_{sc}$, the above spectrum is always fully gapped except for a single gap closing point at $k=0$ for\footnote{Recall that we are assuming $\alpha>0$ throughout this entire section. In the absence of SOI, i.e., for $\alpha=0$, the bulk gap can also close at points other than $k=0$ and the arguments presented in the following do no longer hold.}
\begin{equation}
\Delta_Z^2=\mu^2+\Delta_{sc}^2.\label{eq:gap_closing}
\end{equation}
As we will see in the next subsection, this closing and reopening of the bulk gap corresponds to a topological phase transition between a trivial phase and a topologically nontrivial phase characterized by the emergence of an MBS at each wire end.

\begin{figure*}[t]
	\centering
	\includegraphics[width=\textwidth]{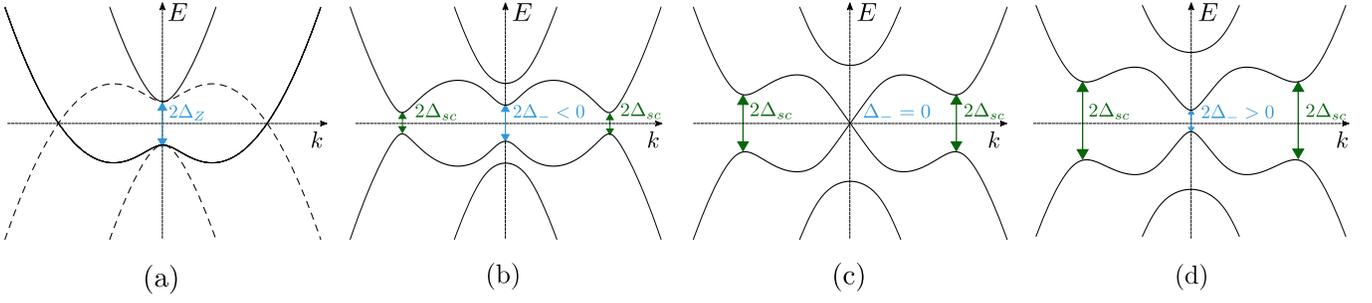}
	\caption{Bulk spectrum of a Rashba nanowire in the presence of a magnetic field and proximity-induced superconductivity, see Eq.~(\ref{eq:spectrum}). For simplicity, we depict the case $\mu=0$. (a) In the absence of superconductivity, i.e., $\Delta_{sc}=0$, a Zeeman gap of size $2\Delta_Z$ is opened around $k=0$, while the exterior branches remain gapless. (b) If a small but finite $\Delta_{sc}>0$ is turned on, the interior gap is modified to  $2\Delta_-$ with $\Delta_-<0$, while a gap of size $2\Delta_{sc}$ opens around $k=\pm 2k_{so}$. (c) As $\Delta_{sc}$ is increased, we eventually reach $\Delta_Z=\Delta_{sc}$, where the interior gap closes. (d) For $\Delta_{sc}>\Delta_Z$, the interior gap reopens with $\Delta_->0$. The closing and reopening of the bulk gap shown in panels (b)-(d) marks the phase transition between the topologically nontrivial phase with $\Delta_-<0$ and the trivial phase with $\Delta_->0$.}
	\label{fig:spectrum}
\end{figure*}

\subsection{Majorana wave functions}
\label{subsec:MBSs_Rashba}

In the following, we focus on the regime of strong SOI with $0\leq \Delta_Z,\Delta_{sc}\ll E_{so}$. If we furthermore assume $|\mu|\ll E_{so}$, we can linearize the spectrum around the Fermi points for $\mu=0$, which are given by $k=0$ and $k=\pm 2k_{so}$ (see again Fig.~\ref{fig:spectrumRashba}). The linearized fields take the form
\begin{align}
\Psi_\uparrow(x)&=e^{-2ik_{so}x}L_\uparrow(x)+R_\uparrow(x),\\
\Psi_{\downarrow}(x)&=L_{\downarrow}(x)+e^{2ik_{so}x}R_{\downarrow}(x),
\end{align}
where $R_\sigma(x)$ [$L_\sigma(x)$] is a slowly varying right-moving [left-moving] field and where we have explicitly taken out the rapidly oscillating factors $e^{\pm 2ik_{so}x}$. In the linearized model, the kinetic term takes the form
\begin{equation}
H_{kin}=-i\hbar v_F\sum_\sigma\int dx\,[R_\sigma^\dagger(x)\partial_x R_\sigma(x)-L_\sigma^\dagger(x)\partial_x L_\sigma(x)],
\end{equation}
where the Fermi velocity is given by $v_F=\alpha/\hbar$ and where we have dropped all rapidly oscillating terms.\footnote{This is justified if the period of oscillation $\propto\pi/(2k_{so})$ is much smaller than all other relevant length scales in the problem, since the corresponding contributions average out in the spatial integral.} 
Similarly, the Zeeman term now reads
\begin{equation}
H_Z=\Delta_Z\int dx\,R_\uparrow^\dagger(x)L_{\downarrow}(x)+\mathrm{H.c.},
\end{equation}
while the superconducting term takes the form
\begin{equation}
H_{sc}=\Delta_{sc}\int dx\,[R_\uparrow(x)L_{\downarrow}(x)+L_\uparrow(x)R_{\downarrow}(x)]+\mathrm{H.c.}
\end{equation}
As an additional simplification, we note that the interior and exterior branches are only coupled among themselves. This allows us to separate the total Hamiltonian into two decoupled subsystems $H_i$ and $H_e$. For $l\in\{i,e\}$ we can write $H_l=\frac{1}{2}\int dx\,\Phi_l^\dagger(x)\mathcal{H}_l(x)\Phi_l(x)$ with $\Phi_i^\dagger=(R_\uparrow^\dagger,L_{\downarrow}^\dagger,R_\uparrow,L_{\downarrow})$ and
\begin{equation}
\mathcal{H}_i(x)=-i\hbar v_F\partial_x\sigma_z-\mu\eta_z+\Delta_Z\eta_z\sigma_x+\Delta_{sc}\eta_y\sigma_y
\end{equation}
for the interior branches and similarly $\Phi_e^\dagger=(L_\uparrow^\dagger,R_{\downarrow}^\dagger,L_\uparrow,R_{\downarrow})$ and
\begin{equation}
\mathcal{H}_e(x)=i\hbar v_F\partial_x\sigma_z-\mu\eta_z+\Delta_{sc}\eta_y\sigma_y
\end{equation}
for the exterior branches. Passing to momentum space once again, the bulk energy spectra for the interior and exterior branches are readily found to be
\begin{align}
E_{i,\pm}^2(k)&=(\hbar v_Fk)^2+(\Delta_+^2+\Delta_-^2)/2\nonumber\\&\quad\pm 2\sqrt{[(\Delta_+^2-\Delta_-^2)/4]^2+(\hbar v_Fk)^2\mu^2},\label{eq:spectrum_interior}\\
E_{e,\pm}^2(k)&=(\hbar v_Fk\pm\mu)^2+\Delta_{sc}^2,\label{eq:spectrum_exterior}
\end{align}
where the momentum $k$ is now taken from the Fermi points and where we have defined $\Delta_\pm=\sqrt{\mu^2+\Delta_{sc}^2}\pm\Delta_{Z}$. Consistent with Eq.~(\ref{eq:gap_closing}), we find from Eq.~(\ref{eq:spectrum_interior}) that the gap for the interior branches closes at $k=0$ when $\Delta_-=0$, while the exterior branches are always fully gapped for any finite $\Delta_{sc}$. 

Figure~\ref{fig:spectrum} illustrates this gap closing and reopening in more detail, where we chose to depict the case $\mu=0$ for simplicity. In this case, we obtain the particularly simple expressions $\Delta_\pm=\Delta_{sc}\pm\Delta_Z$. Let us first consider the case $\Delta_{sc}=0$, see Fig.~\ref{fig:spectrum}(a). In this case, a gap of size $2\Delta_Z$ is opened for the interior branches, while the exterior branches stay gapless. We are thus in a so-called \emph{helical} regime where there is only one gapless right-moving bulk mode with spin down and one left-moving bulk mode with spin up.~\cite{Streda2003,Klinovaja2011} If one now turns on a small but finite $\Delta_{sc}>0$, the interior gap is modified to $2\Delta_-$ with $\Delta_-<0$, while the exterior branches open a gap of size $2\Delta_{sc}$, see Fig.~\ref{fig:spectrum}(b). As $\Delta_{sc}$ is increased, we eventually reach $\Delta_Z=\Delta_{sc}$, where the interior gap closes, see  Fig.~\ref{fig:spectrum}(c). Finally, for $\Delta_{sc}>\Delta_Z$, the interior gap reopens with $\Delta_->0$, see Fig.~\ref{fig:spectrum}(d).

From our preliminary knowledge of the Kitaev chain, we can already guess that the above closing and reopening of the bulk gap could mark the phase transition between a topologically nontrivial phase with $\Delta_-<0$ and a trivial phase with $\Delta_->0$. Indeed, since the phase with $\Delta_-<0$ exhibits a single pair of helical (and, thus, effectively spinless) bulk modes gapped out by proximity-induced superconductivity, we find ourselves in a similar situation as in the Kitaev chain---with the difference that now we started from a spinful model and a conventional superconductor rather than just assuming spinless electrons with $p$-wave pairing. In the following, we will confirm that the phase with $\Delta_-<0$ is indeed the topological phase by demonstrating that MBSs emerge at the wire ends if and only if $\Delta_-<0$. This is done by explicitly solving the BdG equations corresponding to the linearized model for localized zero-energy bound states, where we again focus on the case $\mu=0$ for simplicity.

As a first step, we determine zero-energy solutions of $H_i$ and $H_e$ independently. These are readily obtained by solving the BdG equation $\mathcal{H}_l\phi_l(x)=0$ for $l\in\{i,e\}$. In the following, we focus on a semi-infinite system with a single edge at $x=0$. In order to obtain normalizable eigenfunctions localized to this edge, we make an Ansatz for an exponentially decaying eigenfunction $\phi_l(x)=\phi_l(0)e^{-x/\xi_l}$, where $\xi_l>0$ is a localization length that remains to be determined. Plugging in this Ansatz and imposing $\mathrm{det}[\mathcal{H}_l]=0$, we find the possible values of $\xi_l$ to be
\begin{align}
\xi_{i}&=\alpha/|\Delta_{sc}-\Delta_Z|,\\ \xi_{i}'&=\alpha/(\Delta_{sc}+\Delta_Z)
\end{align}
for the interior branches and
\begin{equation}
\xi_e=\alpha/\Delta_{sc}
\end{equation}
for the exterior branches. Note again that we assume $\Delta_Z,\Delta_{sc}\geq0$ throughout this entire section, such that the above localization lengths are finite and positive whenever the system is fully gapped. Now solving for the corresponding eigenfunctions, we find two linearly independent, exponentially decaying solutions for the interior and exterior branches each. Up to normalization, these read
\begin{align}
\phi_{i,1}(x)&=\begin{pmatrix}-ip\\1\\ip\\1\end{pmatrix}e^{-x/\xi_i},\ \  \phi_{i,2}(x)=\begin{pmatrix}-i\\-1\\-i\\1\end{pmatrix}e^{-x/\xi_i'},\\
\phi_{e,1}(x)&=\begin{pmatrix}i\\1\\-i\\1\end{pmatrix}e^{-x/\xi_e},\ \ 
\phi_{e,2}(x)=\begin{pmatrix}i\\-1\\i\\1\end{pmatrix}e^{-x/\xi_e},
\end{align}
where we have defined $p=\mathrm{sgn}(\Delta_{sc}-\Delta_Z)$. In second-quantized form, the full zero modes then read $\gamma_{l,s}=\int dx\, \phi_{l,s}^\dagger(x)\Phi_l(x)$ for $s\in\{1,2\}$. The oscillating phase factors $\pm e^{2ik_{so}x}$ can now be reincorporated by going back to the original basis $\Psi^\dagger=(\Psi_\uparrow^\dagger,\Psi_{\downarrow}^\dagger,\Psi_\uparrow,\Psi_{\downarrow })$ and writing $\gamma_{l,s}=\int dx\, \psi_{l,s}^\dagger(x)\Psi(x)$, where we have defined $\psi_{i,1}(x)=\phi_{i,1}(x)$,
$\psi_{i,2}(x)=\phi_{i,2}(x)$ and
\begin{align}
\psi_{e,1}(x)&=\begin{pmatrix}ie^{-2ik_{so}x}\\e^{2ik_{so}x}\\-ie^{2ik_{so}x}\\e^{-2ik_{so}x}\end{pmatrix}e^{-x/\xi_e},\\
\psi_{e,2}(x)&=\begin{pmatrix}ie^{-2ik_{so}x}\\-e^{2ik_{so}x}\\ie^{2ik_{so}x}\\e^{-2ik_{so}x}\end{pmatrix}e^{-x/\xi_e},
\end{align}
and where we again neglect all rapidly oscillating terms.

\begin{figure*}[htb]
	\centering
	\includegraphics[width=1\textwidth]{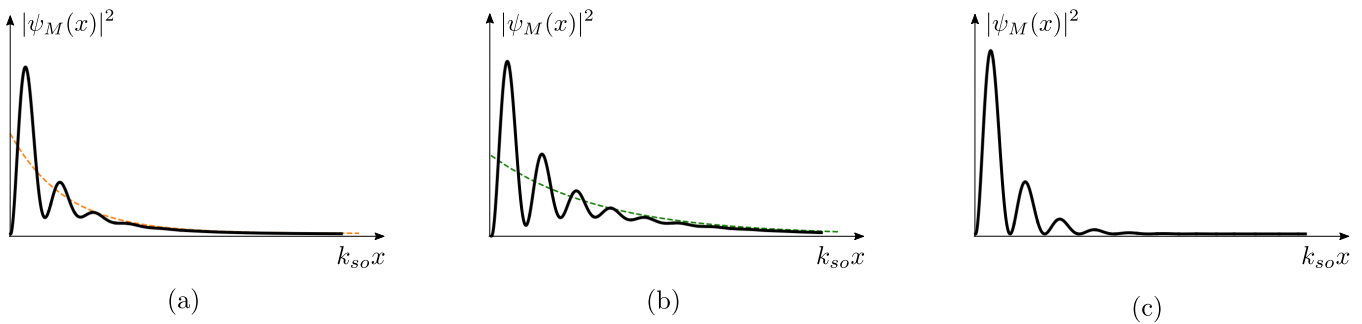}
	\caption{MBS probability density $|\psi_M(x)|^2$ obtained from Eq.~(\ref{eq:Majorana}) for different parameter values. In all cases, $|\psi_M(x)|^2$ exhibits characteristic oscillations with a period of $\pi/k_{so}$ and decays exponentially into the bulk. (a) Deep in the topological phase, there are relatively few oscillations before a uniform exponential decay with decay length $\xi_e$ (orange line) sets in. Here, we used $\Delta_Z/E_{so}=0.5$, $\Delta_{sc}/E_{so}=0.1$. (b) Close to the topological phase transition, the MBS probability density decays more slowly. At long distances, the decay length is given by $\xi_i$ (green line). Here, we used $\Delta_Z/E_{so}=0.3$, $\Delta_{sc}/E_{so}=0.25$. (c) Intermediate regime with $\Delta_Z/E_{so}=0.4$, $\Delta_{sc}/E_{so}=0.2$, where the two decay lengths are equal.}
	\label{fig:wave_functions}
\end{figure*}

As the next and final step, we now turn to the issue of boundary conditions. At the left edge of the system at $x=0$, we demand that our zero-energy wave function $\psi_M(x)\propto\sum_{l,s}c_{l,s}\psi_{l,s}(x)$ satisfies vanishing boundary conditions $\psi_M(x=0)=0$. Here, there is an important difference between the two topologically nonequivalent regimes $\Delta_{sc}>\Delta_Z$ and $\Delta_{sc}<\Delta_Z$. In the first case, the four solutions $\psi_{l,s}$ are linearly independent at $x=0$, showing that the boundary condition can never be fulfilled. This phase corresponds to the topologically trivial one with no zero-energy bound states at the wire ends. In the second case, however, we find that the linear combination $\psi_M(x)\propto\psi_{i,1}(x)-\psi_{e,1}(x)$ satisfies $\psi_M(0)=0$. Explicitly, the total wave function is then given by 
%/4
\begin{equation}
\psi_M(x)=\begin{pmatrix}i\\1\\-i\\1\end{pmatrix}e^{-x/\xi_i}-\begin{pmatrix}ie^{-2ik_{so}x}\\e^{2ik_{so}x}\\-ie^{2ik_{so}x}\\e^{-2ik_{so}x}\end{pmatrix}e^{-x/\xi_e},\label{eq:Majorana}
\end{equation}
where we have suppressed a normalization factor. Furthermore, we can also check that this solution indeed corresponds to a Majorana bound state: The operator $\gamma_M=\int dx\, \psi_M^\dagger(x)\Psi(x)$ satisfies the Majorana property $\gamma_M=\gamma_M^\dagger$ if and only if the wave function satisfies $\psi_M(x)=(f(x),g(x),f^*(x),g^*(x))^T$ for arbitrary functions $f$ and $g$ up to normalization. Our wave function $\psi_M(x)$ given in Eq.~(\ref{eq:Majorana}) can readily be brought into this form by defining $f(x)=ig^*(x)=i(e^{-x/\xi_i}-e^{-2ik_{so}x}e^{-x/\xi_e})$, which confirms that we are indeed looking at an MBS. Equivalently, we can also understand the Majorana property directly from the particle-hole symmetry operation defined in Eq.~(\ref{eq:ph_symm}). Indeed, the above condition on $\psi_M(x)$ is nothing but $U_C\psi^*_M(x)=\psi_M(x)$. This once again reflects the fact that an isolated MBS is `its own partner' under particle-hole symmetry and therefore has to stay pinned to zero energy.

Figure~\ref{fig:wave_functions} shows example plots for the MBS probability density for different values of $\Delta_Z$ and $\Delta_{sc}$. From Eq.~(\ref{eq:Majorana}) we note that there are two possibly different localization lengths $\xi_i$ and $\xi_e$ entering the Majorana wave function, where $\xi_i$ ($\xi_e$) comes from the contribution of the interior (exterior) branches. The interplay between these two contributions causes oscillations in the MBS probability density with a period of $\pi/k_{so}$.

Note that while above we have focused on the case $\mu=0$ for analytical simplicity, MBSs also exist in the more general case of finite $\mu$ as long as the bulk gap remains open. This stability can again be understood from the particle-hole symmetry of the BdG spectrum as discussed in Sec.~\ref{sec:prelim}. Alternatively, the presence of MBSs can also explicitly be verified by, e.g., numerical exact diagonalization of a corresponding tight-binding model. In conclusion, we identify the regime
\begin{equation}
|\Delta_Z|>\sqrt{\mu^2+\Delta_{sc}^2}
\label{eq:top_criterion}
\end{equation}
as the topologically nontrivial phase of the Rashba nanowire model. We note that the same topological criterion could also be obtained directly from the exact bulk spectrum given in Eq.~(22).

Let us close this subsection with a few remarks on the approximations that were made in order to arrive at Eqs.~(\ref{eq:Majorana}). Firstly, we were working in the limit of strong SOI such that the superconducting and Zeeman terms can be treated as weak perturbations to the kinetic Hamiltonian. However, analytical solutions for the MBSs can also be obtained in the opposite limit of weak SOI, see Ref.~\onlinecite{Klinovaja2012}. Secondly, we have focused on a semi-infinite system with a single edge at $x=0$. This corresponds to the ideal case where the MBS at $x=0$ is completely independent of the second MBS at the other end of the system. In a nanowire of finite length, on the other hand, the two MBSs necessarily overlap and hybridize into a fermionic in-gap state with an energy that decreases exponentially with the length of the system. As such, the semi-infinite approximation is justified for large systems. For the explicit fermionic in-gap solutions and the `quasi-MBS' wave functions in a system of arbitrary finite length, we refer the reader to Ref.~\onlinecite{Chua2020}. Thirdly, we have also neglected orbital effects\cite{Lim2013,Nijholt2016,Dmytruk2018,Wojcik2018} caused by the magnetic field. If taken into account,  in the limit of strong SOI, there are regimes in which the amplitude of the oscillating MBS splitting stays constant or even decays with increasing magnetic field, in stark contrast to the commonly studied case where orbital effects of the magnetic field are neglected.\cite{Dmytruk2018}
Last but not least, electron-electron interactions were completely neglected in our considerations. As was shown in Refs.~\onlinecite{Gangadharaiah2011,Stoudenmire2011,Gergs2016,Dominguez2017,Ptok2019,Katsura2015}, MBSs can also survive in the presence of weak to moderate electron-electron interactions. Furthermore, also models with long-range hoppings and long-range pairing interactions have been studied.\cite{Viyuela2016,Cats2018}

\subsection{Rotating magnetic field and synthetic SOI}
\label{subsec:syntheticSOI}

The setup described in the previous subsections requires SOI of Rashba type as a necessary ingredient. In addition, the chemical potential has to be fine-tuned to lie sufficiently close to the spin-orbit energy. These two requirements limit the experimental feasibility of the Rashba nanowire setup, in particular since the intrinsic SOI of the nanowire is a material-dependent property that cannot be fully controlled from the outset. A promising alternative is the so-called \emph{synthetic} SOI induced by a rotating magnetic field generated by, e.g., suitably arranged nanomagnets.\cite{Klinovaja2012b,Kjaergaard2012,Klinovaja2013a,Klinovaja2013b,Matos2017,Desjardins2019,Fatin2016,Zhou2019,Mohanta2019} 
Indeed, a local gauge transformation relates a Rashba nanowire subjected to a uniform magnetic field to a nanowire without SOI subjected to a helical magnetic field.\cite{Braunecker2010} Therefore, both setups exhibit a topologically non-trivial phase with MBSs at the wire ends.

To make this statement explicit, let us consider the spin-dependent gauge transformation
\begin{equation}
\Psi_\sigma(x)=e^{-i\sigma k_{so} x}\tilde\Psi_\sigma(x),
\end{equation}
where the fields $\tilde\Psi_\sigma(x)$ are now defined in a rotating frame. In the rotating frame, $H_0$ defined previously [see Eqs.~(\ref{eq:Hkin}) and (\ref{eq:Hsoi})] takes the form
\begin{equation}
H_{0}=\sum_{\sigma}\int dx\,\tilde\Psi_\sigma^\dagger(x)\left[-\frac{\hbar^2\partial_x^2}{2m}-(\mu+E_{so})\right]\tilde\Psi_\sigma(x).
\end{equation}
This means that the SOI is now absent and the bulk spectrum is effectively given by two identical parabolas---one for spin up and one for spin down---centered around $k=0$. While the superconducting term retains its form in terms of the new fields, the Zeeman term now reads
\begin{equation}
H_Z=\Delta_Z\int dx\,e^{2ik_{so}x}\tilde\Psi_\uparrow^\dagger(x)\tilde\Psi_{\downarrow}(x)+\mathrm{H.c.},
\end{equation}
which takes the form of a Zeeman term induced by a helical magnetic field with a pitch of $2k_{so}$,
\begin{equation}
\tilde B(x)=B(\mathrm{cos}(2k_{so}x),-\mathrm{sin}(2k_{so}x),0).
\end{equation}
As before, MBSs can emerge from this setup  if the chemical potential is tuned to lie in the gap opened by the helical magnetic field. Even more interestingly, however, it has been shown that in certain setups the need to fine-tune the chemical potential is eliminated. This can, for example, be realized in systems with Ruderman-Kittel-Kasuya-Yosida (RKKY) interaction.\cite{Ruderman1954,Kasuya1956,Yosida1957} Indeed, if magnetic impurities are placed on a superconducting substrate, a strong indirect exchange interaction of RKKY-type promotes a helical spin ordering with pitch $2k_F$.\cite{Klinovaja2013c}

\section{Majorana bound states in TI heterostructures}
\label{sec:TI}

We have seen in the previous section that the presence of helical modes is crucial for the emergence of MBSs in 1D TSCs. While in Rashba nanowires the helical regime is realized due to strong SOI in combination with a magnetic field, a related approach---originally proposed in Refs.~\onlinecite{Fu2008,Fu2009} even before the nanowire setup---instead exploits the helical edge states of a 2D TI to engineer MBSs. As opposed to the nanowire case, where time-reversal symmetry needs to be explicitly broken in order to reach the helical regime, the helical regime in TIs emerges in the presence of time-reversal symmetry. Indeed, the key feature of a 2D TI is the existence of a pair of gapless helical edge states with an approximately linear dispersion [see also Fig.~\ref{fig:domain_wall_MBS}(a)] in addition to a fully gapped bulk. The two counterpropagating edge states carry opposite spin projections and are related to each other by time-reversal symmetry.

In the following, we will first review the general mechanism leading to the emergence of bound states at mass domain walls in a system with two counterpropagating linearly dispersing states.~\cite{Jackiw1976,Jackiw1981} Subsequently, we will demonstrate how this can lead to the emergence of MBSs when the edge of a TI is proximitized by a conventional superconductor. As in the Rashba nanowire case, we identify the conditions under which MBSs can emerge in such a system and explicitly obtain their wave function solutions.

We note that we will not give a detailed review of the physics of TIs in this Tutorial. Instead, we work with a basic and rather generic edge-state picture that will be sufficient to understand the emergence of MBSs in such systems. We refer the reader to Refs.~\onlinecite{Hasan2010,Qi2011,Bernevig2013,Asboth2016,Ando2013} for a pedagogical introduction to the field of TIs.

\subsection{Primer: Mass domain walls and Jackiw-Rebbi bound states}
\label{sec:TI_JackiwRebbi}

Let us start by considering a massive 1D Dirac Hamiltonian $H=\int dx\, \Phi^\dagger(x)\mathcal{H}(x)\Phi(x)$ with $\Phi^\dagger=(R^\dagger,L^\dagger)$ and
\begin{equation}
\mathcal{H}(x)=-i\hbar v_F\sigma_z\partial_x-m\sigma_x,
\label{eq:JackiwRebbi}
\end{equation}
where $\sigma_i$ for $i\in \{x,y,z\}$ are Pauli matrices acting in right-/left-mover space. The bulk spectrum of this system is given by $E_\pm(k)=\pm\sqrt{(\hbar v_Fk)^2+m^2}$ and thus is gapped whenever $m\neq 0$. We will now allow the mass term to become position-dependent, i.e., $m=m(x)$. In particular, we consider the effect of a mass domain wall where $m(x\rightarrow-\infty)<0$ and $m(x\rightarrow+\infty)>0$. As was first shown by Jackiw and Rebbi in Ref.~\onlinecite{Jackiw1976}, such a domain wall binds a localized zero-energy state.

\begin{figure}[tb]
	\centering
	\includegraphics[width=0.6\columnwidth]{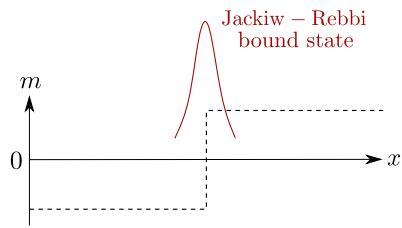}
	\caption{A Dirac Hamiltonian with a mass domain wall hosts a zero-energy bound state localized to the region where the mass changes its sign. As a particularly simple example, we depict the case of a step-function mass term $m(x)=m_0\,\mathrm{sgn}(x)$.}
	\label{fig:domain_wall_JR}
\end{figure} 

Let us illustrate this statement by considering the particularly simple case where the mass has a step-function profile $m(x)=m_0\,\mathrm{sgn}(x)$ with $m_0>0$, see Fig.~\ref{fig:domain_wall_JR}. We can now readily solve for a normalizable zero-energy state by making the Ansatz
\begin{equation}
\phi(x)=\begin{cases}\phi_0^-\,e^{x/\xi_-}&x<0,\\\phi_0^+\,e^{-x/\xi_+}&x>0,\end{cases}
\end{equation}
with the respective localization lengths $\xi_\pm>0$. By solving for zero-energy eigenstates in the intervals $(-\infty,0)$ and $(0,+\infty)$ independently and then imposing continuity of the solution at $x=0$ by requiring $\phi_0^+=\phi_0^-=:\phi_0$, we find that there indeed exists a normalizable zero-energy solution with $\xi_\pm=\hbar v_F/m_0$ and $\phi_0=(1,i)^T$ up to an overall normalization constant. 

While we have presented a concrete example for illustrative purposes, it can be shown that the existence of the above solution does not depend on the exact form of the domain wall. Indeed, for a more general domain-wall profile with the asymptotic behavior $m(x\rightarrow-\infty)<0$ and $m(x\rightarrow+\infty)>0$, we find a zero-energy bound state of the form
\begin{equation}
\phi(x)\propto e^{-\int_0^{x}m(x')dx'/\hbar v_F}\begin{pmatrix}1\\i\end{pmatrix}.
\label{eq:JR_general}
\end{equation}
The fact that the above zero-energy solution exists independently of the exact mass profile function can be understood from the theory of symmetry-protected topological phases of matter. Indeed, the Hamiltonian given in Eq.~(\ref{eq:JackiwRebbi}) has a chiral symmetry\cite{Ryu2010,Chiu2016} expressed by $\{\mathcal{H}(k),\sigma_y\}=0$. The two phases with $m_0\gtrless0$ correspond to two topologically distinct phases that cannot smoothly be deformed into each other unless chiral symmetry is broken. A domain wall between these two distinct phases then hosts a bound state with an energy that is pinned to zero. This follows because chiral symmetry imposes that every state with energy $+E$ has a partner with energy $-E$. The single bound state at the domain wall can therefore not be removed from zero energy unless chiral symmetry is broken. Note that while this argument is superficially similar to the case of particle-hole symmetry discussed earlier, the zero-energy state found in the Jackiw-Rebbi model is an ordinary fermionic zero-mode and not an MBS. This can easily be seen from Eq.~(48). Generally, since we are dealing with a usual single-particle Hamiltonian rather than a BdG Hamiltonian, there is no redundancy in the spectrum of excitations and each eigenstate of the Hamiltonian corresponds to a usual fermionic excitation.

\begin{figure*}[tb]
	\centering
	\includegraphics[width=\textwidth]{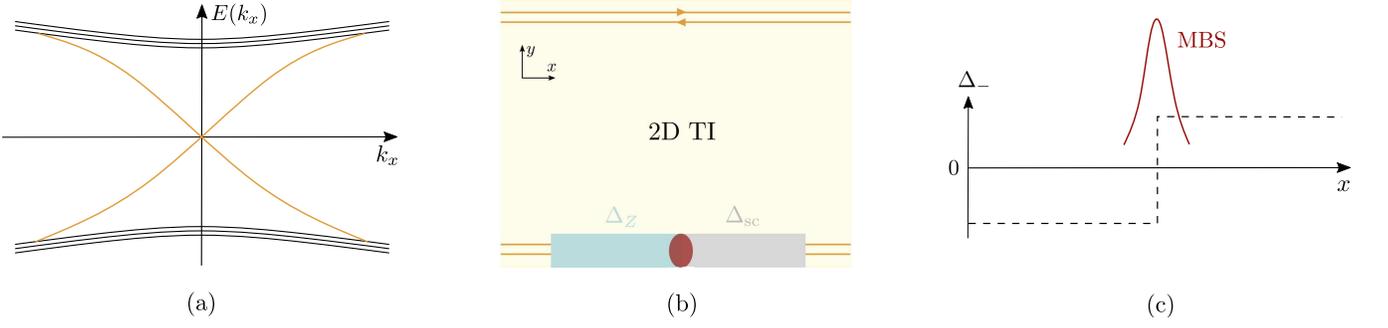}
	\caption{(a) Schematic spectrum of a 2D TI. We take the sample to be finite along the $y$ axis and infinite along the $x$ axis such that $k_x$ remains a good quantum number. While the bulk exhibits an energy gap (black lines), there is a pair of gapless helical edge states (orange lines) propagating along the edges of the sample. (b) The gapless helical edge states of a 2D TI can be gapped out either by proximity-induced superconductivity (grey region) or by a magnetic term (light blue region). An interface between these two opposite gap-opening mechanisms hosts an MBS (shown in red). (c) The existence of a bound state can be understood from the fact that the Dirac mass $\Delta_-=\Delta_{sc}-\Delta_Z$ changes sign at the interface between the two regions. In that sense, the emerging MBS can be interpreted as a special variant of a Jackiw-Rebbi bound state (see Fig.~\ref{fig:domain_wall_JR}) protected by particle-hole symmetry rather than chiral symmetry.}
	\label{fig:domain_wall_MBS}
\end{figure*} 

\subsection{MBSs at domain walls in 2D TIs}
\label{sec:TI_domain_wall}

With a basic understanding of the Jackiw-Rebbi model and the formation of bound states at mass domain walls, let us turn our attention to the construction of MBSs from the edge states of a 2D TI. For simplicity, we focus on the case where the spin component along one axis, say, the $z$ axis, is conserved. Deep inside the bulk gap, we can linearize the edge state dispersion around the Dirac point $k=0$, see Fig.~\ref{fig:domain_wall_MBS}(a). Therefore, the effective low-energy Hamiltonian describing the 1D edge states can be written in terms of a left-moving field $L(x)$ with spin up and a right-moving field $R(x)$ with spin down. The kinetic term then takes the form of a Dirac Hamiltonian
\begin{equation}
H_{kin}=-i\hbar v_F\int dx\,[R^\dagger(x)\partial_x R(x)-L^\dagger(x)\partial_x L(x)].
\end{equation}
We will now consider two different gap-opening mechanisms for the helical edge states: Firstly, an in-plane Zeeman term breaks time-reversal symmetry and couples the left- and the right-moving edge states due to the fact that they carry opposite spin projections. The corresponding term in the Hamiltonian can be written as
\begin{equation}
H_Z=\Delta_Z\int dx\, R^\dagger(x)L(x)+\mathrm{H.c.},
\end{equation}
where $\Delta_Z$ is the strength of the Zeeman term. Secondly, placing the TI edge in proximity to a conventional $s$-wave superconductor gives rise to a proximity-induced superconducting term described by
\begin{equation}
H_{sc}=\Delta_{sc}\int dx\, R(x)L(x)+\mathrm{H.c.},
\end{equation}
where the superconducting pairing potential $\Delta_{sc}$ is taken to be real. If we additionally allow for a small but possibly finite chemical potential (measured from the Dirac point), it is easy to check that the bulk spectrum of the total Hamiltonian $H=H_{kin}+H_Z+H_{sc}$ is once again given by Eq.~(\ref{eq:spectrum_interior}). As such, we already know that the bulk will be fully gapped unless
\begin{equation}
\Delta_Z^2=\Delta_{sc}^2+\mu^2.
\end{equation}
Motivated by our previous discussion of the Jackiw-Rebbi model, we would like to see what happens at a domain wall between the two topologically inequivalent phases separated by this gap closing point. If we assume for concreteness that $\Delta_Z\geq0$, the two inequivalent gapped phases are characterized by $\Delta_-\gtrless 0$, where we have defined $\Delta_-=\sqrt{\Delta_{sc}^2+\mu^2}-\Delta_Z$ in the exact same way as in Sec.~\ref{sec:nanowires}. %For $\mu=0$, one can check that by rewriting the Hamiltonian in a Majorana basis via the unitary transformation $U=e^{i\pi\eta_y\sigma_z/4}$, it can be mapped onto two copies of Eq.~(\ref{eq:JackiwRebbi}) with Dirac masses $\Delta_\pm$. Consequently, we expect to find a zero-energy bound state at a domain wall where one of the Dirac mass changes sign. Indeed, this bound state turns out to be an MBS protected by particle-hole symmetry.
We will now confirm the existence of a domain wall bound state by an explicit calculation. Indeed, it will turn out that this domain wall bound state is an MBS protected by particle-hole symmetry. In the following, we set $\mu=0$ for simplicity. Let us consider a TI edge separated into two segments, where one segment is in contact to a ferromagnet and the other one is proximitized by an $s$-wave superconductor, see Fig.~\ref{fig:domain_wall_MBS}(b). For simplicity, we will take the two segments to be semi-infinite and assume a step-function profile for $\Delta_Z,\Delta_{sc}>0$. The total Hamiltonian incorporating the domain wall then reads $H=\frac{1}{2}\int dx\, \Phi^\dagger(x)\mathcal{H}(x)\Phi(x)$ with $\Phi^\dagger=(R^\dagger,L^\dagger,R,L)$ and
\begin{equation}
\mathcal{H}(x)=-i\hbar v_F\partial_x\sigma_z+\Theta(-x)\Delta_Z\eta_z\sigma_x+\Theta(x)\Delta_{sc}\eta_y\sigma_y,
\end{equation}
where $\sigma_i$ ($\eta_i$) for $i\in\{x,y,z\}$ are Pauli matrices acting in right-/left-mover (particle-hole) space and $\Theta(x)$ is the Heaviside step function. We can now adopt the strategy developed in the previous subsection to show that there is a single MBS localized at the interface, which we have taken to lie at $x=0$ without loss of generality. This involves first solving for decaying eigenstates for $x>0$ and $x<0$ separately and then forming suitable linear combinations to satisfy the boundary condition at the interface. With the Ansatz $\phi_+(x)=\phi_+(0)e^{-x/\xi_+}$  for $x>0$ [$\phi_-(x)=\phi_-(0)e^{x/\xi_-}$  for $x<0$], we immediately obtain the localization lengths
\begin{align}
\xi_-&=\hbar v_F/\Delta_Z,\quad x<0,\\
\xi_+&=\hbar v_F/\Delta_{sc},\quad x>0.
\end{align}
For each of the two segments, we find that the subspace of exponentially decaying zero-energy eigenfunctions is two-fold degenerate. Explicitly, this subspace is spanned by the linearly independent eigenfunctions
\begin{align}
\phi_{-,1}(x)&=\begin{pmatrix}-i\\1\\0\\0\end{pmatrix}e^{x/\xi_-},\quad\phi_{-,2}(x)=\begin{pmatrix}0\\0\\i\\1\end{pmatrix}e^{x/\xi_-},\\
\phi_{+,1}(x)&=\begin{pmatrix}-i\\0\\0\\1\end{pmatrix}e^{-x/\xi_+},\quad\phi_{+,2}(x)=\begin{pmatrix}0\\-i\\1\\0\end{pmatrix}e^{-x/\xi_+},
\end{align}
where we suppress a normalization factor. Finally, demanding that the wave function is continuous at $x=0$, we obtain a single bound state solution
\begin{equation}
\phi_M(x)=\begin{pmatrix}-i\\1\\i\\1\end{pmatrix}\left[\Theta(-x)e^{x/\xi_-}+\Theta(x)e^{-x/\xi_+}\right],
\end{equation}
where we have again suppressed a normalization factor. We can readily verify that this solution does indeed satisfy the Majorana property as it can be brought into the form $\phi_M(x)=(f(x),g(x),f^*(x),g^*(x))^T$ with
\begin{equation}
	f(x)=-ig(x)=\begin{cases}
	-ie^{x/\xi_-},&x<0,\\-ie^{-x/\xi_+},&x>0.\end{cases}
\end{equation}
Again, the presence of the MBS does not depend on the exact profile of the domain wall. Indeed, any interface between a region dominated by magnetic field ($\Delta_-<0$) and a region dominated by superconductivity ($\Delta_->0$) will host an MBS, see Fig.~\ref{fig:domain_wall_MBS}(c). This reflects the fact that the two phases with $\Delta_-\gtrless0 $ are topologically distinct and cannot be connected to each other without closing and reopening the bulk gap. Similarly, the MBS will also persist in the presence of a finite $\mu$ as long as the bulk gap remains open. Motivated by the basic mechanism discussed above, various realizations of MBSs in TI heterostructures have been proposed theoretically.\cite{Cheng2012,Motruk2013,Klinovaja2014b,Klinovaja2015,Schrade2015,Li2016,Ziani2020,Fleckenstein2021}

While the ideas discussed in this section are extremely appealing from a conceptual point of view, significant challenges are met in their experimental realization. This is mainly due to the fact that 2D TIs are highly nontrivial topological materials, the experimental detection and manipulation of which requires significant effort. Nevertheless, recent works report signatures of proximity-induced superconductivity on TI edge states.\cite{Bocquillon2017,Sun2017} Alternatively, the TI edge states could also be replaced by helical hinge states of a three-dimensional (3D) second-order TI.\footnote{As opposed to conventional 3D TIs with gapless surface states, 3D \emph{second-order} TIs have gapped surfaces but host gapless chiral or helical 1D modes propagating along the hinges of the sample,\cite{Schindler2018,Langbehn2017,Song2017,Benalcazar2017b} see also Ref.~\onlinecite{Parameswaran2017} for a popular summary.} In this latter case, zero-bias peaks consistent with MBSs were measured at domain walls between ferromagnetic and superconducting domains.\cite{Jack2019} 

In variations of the above setup, MBSs could also be engineered from quantum Hall edge states in a suitable sample geometry.\cite{Lindner2012,Clarke2013,Lee2017} As a side remark, we mention that proximitized quantum Hall systems have raised significant interest not only as potential hosts for MBSs, but also for \emph{propagating} chiral Majorana edge states.\cite{Qi2010} Signatures of proximity-induced superconductivity in quantum Hall edge states were studied in a series of recent works.\cite{Hart2014,Amet2016,Lee2017,Draelos2018,Zhao2020,Gul2020}

Finally, it is worth noting that the above setups have also raised significant interest due to their possible generalization to the regime of strong electron-electron interactions, where even more exotic bound states are predicted to emerge. In particular, if the TI (quantum Hall) edge states are replaced by \emph{fractional} TI (quantum Hall) edge states, domain walls between competing gap-opening mechanisms are theoretically predicted to host \emph{parafermion} bound states.~\cite{Cheng2012,Clarke2013,Lindner2012,Motruk2013,Klinovaja2014b,Gul2020} These exotic zero-energy modes can be seen as a formal generalization of MBSs. In particular, while a pair of MBSs encodes a two-fold ground state degeneracy, a pair of $\mathbb{Z}_N$ parafermion zero modes is generally associated with an $N$-fold topologically protected ground state degeneracy. Braiding operations of parafermion zero modes then realize an even richer set of non-Abelian rotations on this ground state manifold. We refer the interested reader to Refs.~\onlinecite{Alicea2016,Alicea2015,Schmidt2020} for pedagogical reviews of the topic.

On a related note, let us mention that systems of interacting MBSs provide an exciting playground to study novel phases of matter with even more exotic properties such as topological order, emergent supersymmetry, or chaotic behavior related to the Sachdev-Ye-Kitaev (SYK) model,\cite{Rahmani2015,Hassler2012,Kells2014,Terhal2012,Chiu2015b,Chew2017} see also Ref.~\onlinecite{Rahmani2019} for a review.

\section{Alternative realizations of Majorana bound states in 1D systems}
\label{sec:others}

The above ideas have inspired an ever-growing list of proposals aimed at the realization of MBSs in topologically non-trivial 1D systems. In the following, we will briefly highlight several directions that we consider to be of particularly high importance, referring the reader to the excellent reviews Refs.~\onlinecite{Alicea2012,Beenakker2013,Aguado2017,Lutchyn2018,Stanescu2013,DasSarma2015,Sato2017,Sato2016,Leijnse2012,Elliott2015,Pawlak2019,vonOppen2017} for a more exhaustive overview of the existing proposals.

Apart from nanowires, there are several alternative (quasi-)1D platforms that can host MBSs. In particular, extensive research has been carried out on graphene-based structures such as carbon nanoribbons and carbon nanotubes.~\cite{Klinovaja2012c,Egger2012,Sau2013,Dutreix2014,Klinovaja2013a,Marganska2018} Since SOI effects in pristine graphene are generally weak, these setups often rely on synthetic SOI as discussed in Sec.~\ref{subsec:syntheticSOI}. 

Furthermore, a topological superconducting phase with MBSs has also been predicted to occur in quasi-1D nanowires fabricated from three-dimensional TI materials. In particular, when the wire is proximitized by a conventional $s$-wave superconductor and a magnetic field is applied parallel to the wire, well-localized MBSs can emerge at the wire ends both in the presence\cite{Cook2011,Cook2012} or in the absence\cite{Legg2021} of a vortex in the proximity-induced pairing potential. In this latter case, a non-uniform chemical potential in the wire cross-section is responsible for an exceptionally strong effective SOI.

Another family of promising proposals involves chains of magnetic impurities deposited on a superconducting substrate. Again, both a helical magnetic ordering of the impurities or---formally equivalent---a ferromagnetic ordering in combination with strong SOI may lead to the formation of MBSs. If the impurities are placed sufficiently close to each other, the chain can be described as an effective 1D nanowire and MBSs emerge in the same way as discussed in Sec.~\ref{sec:nanowires}.~\cite{Klinovaja2013c,Vazifeh2013,Braunecker2013} For larger inter-impurity distances, on the other hand, a different mechanism becomes important. Indeed, if the exchange interaction between a magnetic atom and the quasiparticles in the superconductor is sufficiently strong, a localized sub-gap Yu-Shiba-Rusinov (YSR) state will emerge.~\cite{Yu1965,Shiba1968,Rusinov1969} In the case of many magnetic impurities, the corresponding YSR states can overlap and form a chain. Such a YSR chain shows many similarities to the Kitaev chain~\cite{Kitaev2001} and, under specific assumptions, can be mapped onto the latter. As such, it becomes clear that a YSR chain can realize a topologically non-trivial phase with MBSs at its ends.~\cite{Choy2011,Nadj-Perge2013,Pientka2013,Li2014,Pientka2014,Heimes2014,Poyhonen2014,Hoffman2016,Andolina2017,Theiler2019} Indeed, a series of experiments reported robust zero-bias peaks in chains of iron (Fe) or cobalt (Co) atoms placed on a superconductor.~\cite{Nadj-Perge2014,Ruby2015,Pawlak2016,Feldman2016,Ruby2017,Jeon2017,Kim2018} For a detailed review of MBSs in magnetic chains we refer the reader to Refs.~\onlinecite{Pawlak2019,Jack2021}. We note that even more complicated non-collinear magnetic textures, such as skyrmions or skyrmion chains, were proposed as an alternative route to generate topological superconducting phases.\cite{Nakosai2013a,Poyhonen2016,Yang2016,Garnier2019,Gungordu2018,Steffensen2020,Rex2020,Kubetzka2020,Mascot2020,Mohanta2020,Diaz2021}

Alternatively, we note that external driving provides a powerful tool to turn initially nontopological materials into topological ones.\cite{Harper2020,Rudner2020} Indeed, it has been shown that a time-dependent magnetic or electric field can give rise to Floquet Majorana fermions.\cite{Kundu2013,Reynoso2013,Thakurathi2013,Thakurathi2014,Benito2014,Lago2015,Thakurathi2017,Liu2019,Plekhanov2019} 

Furthermore, one can also consider unconventional superconductors with $p$-wave and $d$-wave pairings.\cite{Tanaka1995,Sengupta2001,Wehling2014,Ando2015,Sasaki2011,Linder2010,Hu1994,Alidoust2021} Here, one should mention also odd-frequency superconductivity.\cite{Linder2019,Tanaka2007,Tanaka2007b,Triola2020,Tanaka2012} The odd-frequency pairing is hugely enhanced at the boundaries of the topological systems hosting MBSs.\cite{Tanaka2012,Ebisu2016,Cayao2018,Fleckenstein2018,Krieger2020}

Last but not least, it is worth mentioning that a magnetic field is not a necessary ingredient to generate MBSs. Indeed, there are many alternative ways in which competing gap-opening mechanisms can realize a phase transition between a trivial and a topological superconducting phase with MBSs. This is of particular importance since strong magnetic fields have a detrimental effect on superconductivity. Efforts to avoid this obstacle have resulted in an increased interest in time-reversal invariant systems. In particular, it has been found that a spinful time-reversal invariant 1D TSC hosts a \emph{Kramers pair} of MBSs at each end in the topologically non-trivial phase.~\cite{Keselman2013,Klinovaja2014,Haim2014,Gaidamaiskas2014,Dumitrescu2014,Haim2016,Ebisu2016,Schrade2017,Thakurathi2018,Aligia2018,Wong2012,Zhang2013,Nakosai2013} Even though localized at the same position in real space, the two MBSs at a given end of the system are then protected from hybridizing by time-reversal symmetry.

\begin{figure*}[tb]
	\centering
	\includegraphics[width=0.85\textwidth]{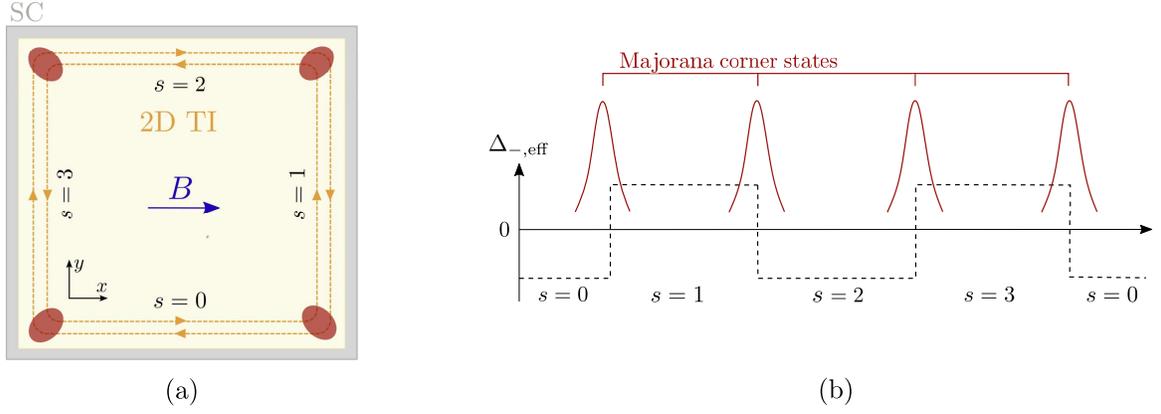}
	\caption{(a) A 2D TI (yellow) described by the Hamiltonian given in Eq.~(\ref{eq:BHZ}) is placed on top of an $s$-wave superconductor (grey) and a Zeeman field is applied along the $x$ direction (blue arrow). The Zeeman and superconducting terms gap out the helical edge states (orange) running along the edges of the sample. If the strength of the Zeeman term is larger than the proximity-induced superconductivity, we find one Majorana corner state (red) at each corner of the sample. (b) The existence of Majorana corner states can be explained by looking at the effective gap $\Delta_{-,\mathrm{eff}}=\Delta_{sc,\mathrm{eff}}-\Delta_{Z,\mathrm{eff}}$, which is obtained by projecting the respective gap-opening terms onto the low-energy subspace spanned by the helical edge states propagating along the edges $s=0,...,3$ of the sample. At each of the four corners,  $\Delta_{-,\mathrm{eff}}$ changes sign, giving rise to a Majorana corner state at each corner.}
	\label{fig:HOTSC_4}
\end{figure*} 

\section{Majorana corner states}
\label{sec:MCSs}

In Sec.~\ref{sec:TI}, a domain wall hosting an MBS was realized through couplings $\Delta_Z(x)$, $\Delta_{sc}(x)$ with an explicit spatial dependence. More recently, the concept of \emph{higher-order} topological insulators and superconductors~\cite{Benalcazar2017,Benalcazar2017b,Geier2018,Peng2017,Schindler2018,Song2017,Imhof2017,Schindler2020} has opened up an alternative avenue towards the realization of MBSs in TI heterostructures. While conventional $d$-dimensional TIs and TSCs exhibit gapless edge states at their $(d-1)$-dimensional boundaries, $n$th-order $d$-dimensional TIs or TSCs exhibit gapless edge states at their $(d-n)$-dimensional boundaries. In particular, a 2D second-order topological superconductor (SOTSC) hosts MBSs at the corners of a rectangular sample.~\cite{Khalaf2018,Zhu2018,Wang2018,Liu2018,Phong2017,Volpez2019,Laubscher2019,Franca2019,Yan2019,Zhang2019,Zhu2019,Wu2019,Wu2019b,Ahn2020,Laubscher2020,Zhang2020,Laubscher2020b}

One particular way to obtain such Majorana corner states is to start from a conventional (first-order) TI or TSC with helical edge states, which are then perturbatively gapped out by small additional terms. Depending on the symmetries of the model as well as the sample geometry, the gap acquired by the edge states is not necessarily of the same size or type for different edges. As such, domain walls between different gap-opening mechanisms emerge naturally even for spatially uniform gap-opening terms. In the following, we will illustrate this concept with two simple examples.

\subsection{SOTSC with four corner states}

Our first example is based on a model introduced in Ref.~\onlinecite{Wu2019b}. We start from a minimal model for a 2D TI, which can be interpreted as a simplified version of the Bernevig-Hughes-Zhang (BHZ) Hamiltonian originally brought forward in Ref.~\onlinecite{Bernevig2006} to describe HgTe quantum wells. We consider a Hamiltonian of the form $H=\sum_\mathbf{k}\Psi_\mathbf{k}^\dagger\mathcal{H}(\mathbf{k})\Psi_\mathbf{k}$ with $\Psi_\mathbf{k}^\dagger=(\psi_{\mathbf{k}\uparrow1}^\dagger$, $\psi_{\mathbf{k}\uparrow\bar{1}}^\dagger$, $\psi_{\mathbf{k}\downarrow1}^\dagger$, $\psi_{\mathbf{k}\downarrow\bar{1}}^\dagger)$, where $\psi_{\mathbf{k}\sigma\tau}^\dagger$ ($\psi_{\mathbf{k}\sigma\tau}$) creates (destroys) an electron with in-plane momentum $\mathbf{k}=(k_x,k_y)$, spin $\sigma$, and an additional local degree of freedom $\tau$. The Hamiltonian density is taken to be
\begin{align}
	\mathcal{H}(\mathbf{k})&=\left(\frac{\hbar^2k_x^2}{2m}+\frac{\hbar^2k_y^2}{2m}+\epsilon\right)\tau_z+\lambda (k_x\sigma_z\tau_x-k_y\tau_y),
	\label{eq:BHZ}
\end{align}
where $\sigma_i$ and $\tau_i$ for $i\in\{x,y,z\}$ are Pauli matrices acting in spin space and on the local degree of freedom $\tau$, respectively. The parameters $m$ and $\lambda$ are model-dependent constants, which we take to be strictly positive in the following. Furthermore, $\epsilon$ describes an energy shift between the two species $\tau\in\{1,\bar1\}$. The above Hamiltonian is time-reversal symmetric as defined in Eq.~(\ref{eq:tr_symm}). Furthermore, the Hamiltonian has a four-fold rotational symmetry
\begin{equation}
\mathcal{H}(k_x,k_y)=U_{\pi/2}\mathcal{H}(-k_y,k_x)U_{\pi/2}^{-1}
\end{equation}
with  $U_{\pi/2}=e^{i\pi \sigma_z(2\tau_0-\tau_z)/4}$. 

For $\epsilon<0$, it is well-known that the above Hamiltonian describes a TI.\cite{Bernevig2006} An explicit calculation confirms that there is indeed a pair of gapless helical edge states propagating along the edges of a finite sample. Let us illustrate this in an example, where we will focus on the edge $s=0$ shown in Fig.~\ref{fig:HOTSC_4}(a). Assuming a semi-infinite geometry such that the system is finite along the $y$ axis and infinite along the $x$ axis, $k_x$ remains a good quantum number, whereas $k_y$ has to be replaced by $-i\partial_y$. We now focus on the simple case $k_x=0$, in which case the Hamiltonian given in Eq.~(\ref{eq:BHZ}) reduces to
\begin{align}
\mathcal{H}(0,-i\partial_y)&=\left(\epsilon-\frac{\hbar^2\partial_y^2}{2m}\right)\tau_z+i\lambda \partial_y\tau_y.
\label{eq:BHZ_reduced}
\end{align}
Solving for normalizable zero-energy solutions $\Phi_\sigma^0(y)$ subject to the boundary condition $\Phi_\sigma^0(0)=0$ now reduces to a standard problem of matching decaying eigenfunctions as discussed in several instances in the previous sections. We readily find
\begin{align}
\Phi^0_{\uparrow}(y)&=(1,1,0,0)^T(e^{-y/\xi_1}-e^{-y/\xi_2}),\\
\Phi^0_{\downarrow}(y)&=(0,0,1,1)^T(e^{-y/\xi_1}-e^{-y/\xi_2}),
\label{eq:solx}
\end{align}
with $\xi_{1/2}=(-\lambda\pm\sqrt{\beta})/(2\epsilon)$ for $\beta=\lambda^2+2\hbar^2\epsilon/m$ and where we have suppressed a normalization factor. One can check that these solutions are indeed exponentially decaying for $y\rightarrow\infty$ if and only if $\epsilon<0$.

Linear contributions in $k_x$ can now in principle be included perturbatively in order to verify that the above solutions do indeed correspond to counterpropagating edge states with opposite velocities.\cite{Wu2019b} However, we will content ourselves with studying the solutions at $k_x=0$ given above. This will allow us to determine the gap that is opened in the edge state spectrum under additional terms that we include perturbatively.

The first of these additional terms is a small in-plane Zeeman field that is taken along the $x$ axis for concreteness, $\mathcal{H}_Z=\Delta_Z\sigma_x$, where we assume $\Delta_Z\geq0$. Exploiting the rotational symmetry of the unperturbed Hamiltonian, we obtain the low-energy projection of the Zeeman term for the edge $s$ via $\langle\Phi^s_{\sigma}|\mathcal{H}_Z|\Phi^s_{\sigma'}\rangle=\langle\Phi^0_{\sigma}|U_{\pi/2}^{-s}\mathcal{H}_ZU_{\pi/2}^{s}|\Phi^0_{\sigma'}\rangle$. Explicitly, this gives us 
\begin{align}
	\langle\Phi^0_{\sigma}|\mathcal{H}_Z|\Phi^0_{\sigma'}\rangle&=-\langle\Phi^2_{\sigma}|\mathcal{H}_Z|\Phi^2_{\sigma'}\rangle=\Delta_Z\delta_{\bar\sigma\sigma'},\\
	\langle\Phi^1_{\sigma}|\mathcal{H}_Z|\Phi^1_{\sigma'}\rangle&=\langle\Phi^3_{\sigma}|\mathcal{H}_Z|\Phi^3_{\sigma'}\rangle=0\quad\forall\,\sigma,\sigma'\label{eq:Zeeman_y}.
\end{align}
As such, the Zeeman term fully gaps out the edge states along the $x$ direction, while the edge states along the $y$ direction are not affected. Furthermore, we consider a superconducting term induced by placing the TI in proximity to a bulk $s$-wave superconductor, see Fig.~\ref{fig:HOTSC_4}(a). The corresponding term in the Hamiltonian is given by $\mathcal{H}_{\mathrm{sc}}=\Delta_{sc}\eta_y\sigma_y$, where $\eta_y$ is an additional Pauli matrix acting in particle-hole space and we assume $\Delta_{sc}$ to be real and non-negative for simplicity. It is clear that superconductivity opens a gap of equal size along all edges of the sample. Therefore, for $\Delta_Z>\Delta_{sc}$, a domain wall of the type discussed in Subsec.~\ref{sec:TI_domain_wall} is naturally realized at all four corners of a rectangular sample. We thus find one Majorana corner state per corner, see Fig.~\ref{fig:HOTSC_4}(b). Apart from the model presented here, other realizations of Majorana corner states via similar mechanisms were proposed in Refs.~\onlinecite{Wang2018,Liu2018,Franca2019,Yan2019,Zhang2019,Zhu2019,Laubscher2020,Wu2019,Zhang2020,Laubscher2020b}.

\begin{figure*}[tb]
	\centering
	\includegraphics[width=0.85\textwidth]{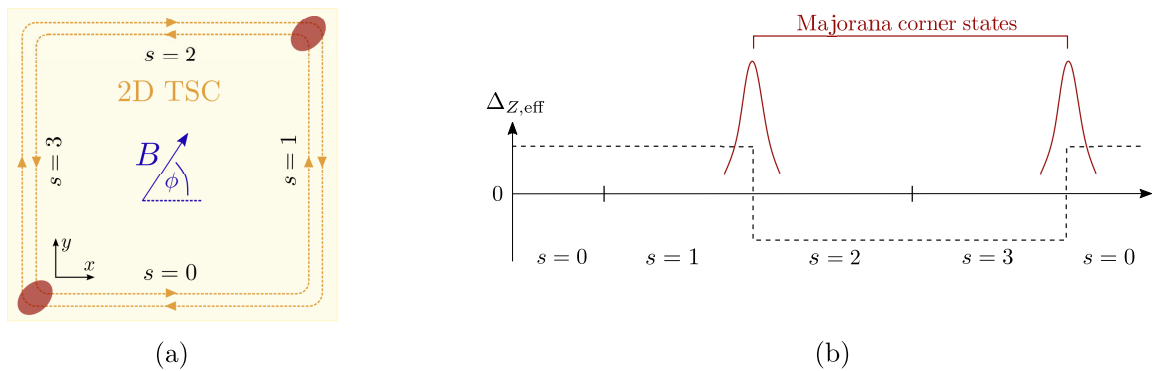}
	\caption{(a) A 2D topological superconductor (yellow) with helical Majorana edge states (orange) described by the Hamiltonian given in Eq.~(\ref{eq:TSC_helical}) is subjected to an in-plane Zeeman field (blue) of an angle $\phi$ with respect to the $x$ axis. The Zeeman field breaks time-reversal symmetry and gaps out the Majorana edge states. MBSs (red) appear at two opposite corners of the sample. (b) The existence of these Majorana corner states can be explained by looking at the effective gap $\Delta_{Z,\mathrm{eff}}$ [see Eq.~(\ref{eq:HZ_eff})], which is obtained by projecting the Zeeman term onto the low-energy subspace spanned by the helical edge states propagating along the edges $s=0,...,3$ of the sample. At two opposite corners of the system, the effective gap changes sign, implying the presence of a bound state of Jackiw-Rebbi type. Since the edge states in the first-order phase correspond to helical Majorana edge states, these zero-energy corner states are MBSs protected by particle-hole symmetry.}
	\label{fig:HOTSC_2}
\end{figure*}

\subsection{SOTSC with two corner states}

Our second example is based on a model introduced in Ref.~\onlinecite{Volpez2019}. We start from a 2D time-reversal invariant topological superconductor with helical Majorana edge states propagating along the edges of a large but finite sample. While a detailed characterization of phases with propagating Majorana edge states is beyond the scope of this Tutorial, we will look at this example from a very simple point of view by just solving for edge state solutions in the usual way and giving an intuitive explanation why these edge states have Majorana character. For a more detailed discussion of phases with propagating Majorana edge states, we refer the reader to several reviews covering this topic.\cite{Alicea2012,Bernevig2013,Aguado2017,Sato2016,Sato2017} 

Explicitly, we consider a Hamiltonian of the form $H=\frac{1}{2}\sum_\mathbf{k}\Psi_\mathbf{k}^\dagger\mathcal{H}(\mathbf{k})\Psi_\mathbf{k}$ with
\begin{align}
	\mathcal{H}(\mathbf{k})&=\left(\frac{\hbar^2k_x^2}{2m}+\frac{\hbar^2k_y^2}{2m}\right)\eta_z+\lambda(k_y\sigma_x-k_x\eta_z\sigma_y)\tau_z\nonumber\\&\quad\ +\Gamma\eta_z\tau_x+\Delta_{sc}\eta_y\tau_z\sigma_y\label{eq:TSC_helical}
\end{align}
and $\Psi_\mathbf{k}^\dagger=(\psi_{\mathbf{k}\uparrow1}^\dagger$, $\psi_{\mathbf{k}\downarrow1}^\dagger$, $\psi_{\mathbf{k}\uparrow\bar{1}}^\dagger$, $\psi_{\mathbf{k}\downarrow\bar{1}}^\dagger,\psi_{-\mathbf{k}\uparrow1}$, $\psi_{-\mathbf{k}\downarrow1}$, $\psi_{-\mathbf{k}\uparrow\bar{1}}$, $\psi_{-\mathbf{k}\downarrow\bar{1}})$. The Pauli matrices $\sigma_i$, $\tau_i$, and $\eta_i$ for $i\in\{x,y,z\}$ have the same meaning as in the previous subsection. The parameters $m$, $\lambda$, and $\Gamma$ depend on the microscopic realization of the model and are taken to be non-negative for simplicity. Furthermore, $\Delta_{sc}$ denotes the strength of the proximity-induced superconducting pairing, which is taken to be real and of opposite sign for the two species $\tau\in\{1,\bar 1\}$. Originally, the above Hamiltonian was introduced in Ref.~\onlinecite{Volpez2018} to describe two tunnel-coupled layers of a 2D electron gas with strong Rashba SOI `sandwiched' between a top and bottom superconductor with a phase difference of $\pi$. In this case, the local degree of freedom $\tau$ corresponds to the layer degree of freedom. The Hamiltonian is particle-hole symmetric and time-reversal symmetric as defined in Eqs.~(\ref{eq:ph_symm}) and (\ref{eq:tr_symm}), respectively. Furthermore, we find a four-fold rotational symmetry 
\begin{equation}
\mathcal{H}(k_x,k_y)=U_{\pi/2}\mathcal{H}(-k_y,k_x)U_{\pi/2}^{-1}
\end{equation}
for $U_{\pi/2}=e^{i\pi\eta_z\sigma_z/4}$.

For $\Gamma>\Delta$, the system realizes a time-reversal invariant topological superconductor, as can be checked by a direct calculation of the edge state wave functions. For this, we focus again on the edge $s=0$ as shown in Fig.~\ref{fig:HOTSC_2}(a). Assuming a semi-infinite geometry, $k_x$ remains a good quantum number, and we restrict ourselves to the simplest case $k_x=0$. After a short calculation outlined in Ref.~\onlinecite{Volpez2018}, we find that the edge state wave functions are in this case given by
\begin{align}
\Phi_{+}^{0}(y)&=(f_1,g_1,f_2,g_2,f^*_1,g^*_1,f^*_2,g^*_2)^T,\\
\Phi_{-}^{0}(y)&=(g_1^*,-f_1^*,g_2^*,-f_2^*,g_1,-f_1,g_2,-f_2)^T,
\end{align}
where $f_1=g_2=-if_2^*=-ig_1^*=e^{-y/\xi}-e^{2ik_{so}y}e^{-y/\xi'}$ with $\xi=\lambda/(\Gamma-\Delta_{sc})$ and $\xi'=\lambda/\Delta_{sc}$. We can now see why these edge states are referred to as propagating Majorana edge states: Indeed, the above solutions at $k_x=0$ satisfy the Majorana property $U_C[\Phi_\pm^0(y)]^*=\Phi_\pm^0(y)$. Again, linear contributions in $k_x$ could now be included perturbatively.\cite{Volpez2019}

If an additional in-plane Zeeman field is added, time-reversal symmetry is broken and the helical edge states are gapped out. The corresponding term in the Hamiltonian can be written as $\mathcal{H}_Z=\Delta_Z[\cos(\phi)\eta_z\sigma_x+\sin(\phi)\sigma_y]$, where the angle $\phi$ describes the orientation of the Zeeman field. It is now straightforward to calculate the projection of the Zeeman term onto the low-energy subspace spanned by the helical edge states for a given edge. We find
\begin{align}
\langle\Phi_+^s|\mathcal{H}_Z|\Phi_-^s\rangle&=\langle\Phi_+^0|U_{\pi/2}^{-s}\mathcal{H}_ZU_{\pi/2}^{s}|\Phi_-^0\rangle\nonumber\\&=i\Delta_Z\cos(\phi-s\pi/2).\label{eq:HZ_eff}
\end{align}
For the edge $s$, this leads to an effective mass term $-\Delta_{Z,\mathrm{eff}}^s\rho_y$ with $\Delta_{Z,\mathrm{eff}}^s=\Delta_Z\cos(\phi-s\pi/2)$ and where $\rho_y$ is a Pauli matrix acting in the low-energy subspace spanned by the helical edge states. We therefore see that the effective mass changes sign at two opposite corners of the sample depending on the in-plane orientation $\phi$ of the magnetic field. Via the Jackiw-Rebbi mechanism discussed in Subsec.~\ref{sec:TI_JackiwRebbi}, these two corners host a zero-energy bound state. Since the helical edge states of the first-order phase already correspond to Majorana edge states, these zero-energy corner states are indeed MBSs protected by particle-hole symmetry. In Fig.~\ref{fig:HOTSC_2}, we illustrate the above mechanism for a generic angle $\phi\in(0,\pi/2)$. Other realizations of Majorana corner states involving similar arguments were proposed in Refs.~\onlinecite{Phong2017,Khalaf2018,Zhu2018,Ahn2020,Laubscher2019}.

\section{Conclusions and Outlook}
\label{sec:conclusions}

In this Tutorial, we have given a pedagogical introduction to the field of MBSs in semiconducting nanostructures. We have reviewed some of the currently most relevant platforms proposed to host MBSs, including proximitized Rashba nanowires in a magnetic field as well as proximitized edge states of topological insulators. In these examples, we have shown how MBSs emerge in the topologically nontrivial phase and how the explicit Majorana wave functions can be obtained by elementary methods. We have discussed the properties of the resulting MBSs and have presented some heuristic arguments about their stability.

Throughout this Tutorial, we chose to focus on a few selected topics that we believe to be of high relevance to experiments while at the same time readily accessible with elementary mathematical tools. As an outlook, let us mention a few relevant aspects of MBSs that were not addressed in this Tutorial.

Firstly, we did not touch upon the characterization of topological superconductors via topological invariants. This important topic is already covered in various existing reviews, see for example Refs.~\onlinecite{Alicea2012,Stanescu2013,Sato2016,Sato2017}.

Secondly, we did not discuss the issues related to the experimental detection of MBSs. Among the standard signatures associated with MBSs is the presence of a robust zero-bias peak of the tunneling conductance as measured, e.g., in transport experiments.\cite{Sengupta2001,Bolech2007,Nilsson2008,Law2009,Flensberg2010,Wimmer2011,Fidkowski2012,Sau2010b,Stanescu2011} However, experiments based solely on zero-bias peaks are nowadays known to be insufficient to conclusively demonstrate the presence of MBSs. Instead, different non-topological states such as, e.g., Andreev bound states can give rise to almost identical zero-bias anomalies. For a detailed review of this problem, we refer to Ref.~\onlinecite{Prada2020}. Although numerous works have reported signatures thought to be unambiguously associated with topological superconductivity, this issue is not yet settled and claims are being reconsidered.\cite{Frolov2021} While the basic theoretical ideas on which the existence of topological superconductors rests are sound and were never disproven, it is their implementation in real materials that poses substantial challenges. In particular, superconductivity in semiconducting structures is typically induced by the proximity effect.\cite{Suominen2017,Hendrickx2018,Bakkers2019,Ridderbos2020,Aggarwal2021} However, if the proximity effect is weak---like in approaches based on sputtering---the superconducting gap is ‘soft’, which is usually attributed to disorder effects. On the other hand, if the hybridization with the bulk superconductor is too strong, the original properties of the underlying semiconductor may be lost. Typically, the superconductor `metallizes' the semiconducting nanostructure, resulting in strongly reduced SOI and $g$-factors.\cite{Reeg2017,Reeg2018,Woods2019,Winkler2019,Kiendl2019} Moreover, due to screening, it is challenging to control the position of the chemical potential in order to tune it to the `sweet spot'. Thus, future experiments need to find ways to avoid or reduce such metallization effects on the semiconductors. This can be achieved, e.g., by adding a thin insulating layer. Another Majorana signature worth mentioning is the fractional Josephson effect arising in a topological superconductor--normal metal--topological superconductor (TS--N--TS) junction, giving rise to an unconventional $4\pi$-periodic Josephson current as opposed to the usual $2\pi$-periodic Josephson current.\cite{Kitaev2001,Fu2009,Jiang2011,Lutchyn2010,Oreg2010,Law2011,Sanjose2012,Ioselevich2011,Dominguez2012,Pikulin2012,Badiane2011} Alternatively, emerging exotic phases could be probed in cavities via photonic transport\cite{Trif2012,Schmidt2013,Dmytruk2015,Cottet2017} characterized by the complex transmission coefficient that relates input and output photonic fields, allowing to probe the system in a global and non-invasive way. Moreover, this could also be used to manipulate states. Generally, these detection methods could be supplemented by additional signatures observable in the bulk such as, e.g., the closing of the topological bulk gap and the inversion of spin polarization in the energy bands.\cite{Gulden2016,Szumniak2017,Chevallier2018,Serina2018,Yang2019,Tamura2019,Sticlet2020,Mashkoori2020} For pedagogical reviews covering experimental Majorana signatures, we refer to Refs.~\onlinecite{Alicea2012,Stanescu2013,Beenakker2013,Elliott2015,Aguado2017}.

Last but not least, we did not discuss the non-Abelian braiding statistics of MBSs in any detail. This choice was motivated by the fact that this Tutorial mainly focuses on 1D systems, where the process of spatially exchanging two MBSs is not as straightforward as in two dimensions. Nevertheless, braiding protocols for MBSs in strictly 1D systems have been brought forward,\cite{Chiu2015,SanJose2016} which, however, typically rely on the presence of additional protecting symmetries or the ability to fine-tune certain system parameters. Alternatively, braiding schemes employing \emph{networks} of topologically nontrivial 1D systems have been proposed,\cite{Alicea2011,Sau2010,Hyart2013,Karzig2016,Aasen2016,Clarke2011,Halperin2012} where the MBSs can either be moved physically or an effective braiding can be realized via tunable couplings between neighboring MBSs at fixed spatial positions. Similarly, also measurement-based braiding schemes allow one to mimic an effective braiding of two MBSs without the need to physically move them.\cite{Bonderson2008,Bonderson2009,Karzig2017} For a review of Majorana braiding statistics and potential applications in topological quantum computation, we refer the interested reader to Refs.~\onlinecite{Alicea2012,Aguado2017,DasSarma2015,Beenakker2013,Leijnse2012,Elliott2015}

\begin{acknowledgments}
We thank Daniel Loss for valuable discussions. This work was supported by the Swiss National Science Foundation and NCCR QSIT. This project received funding from the European Union’s Horizon 2020 research and innovation program (ERC Starting Grant, grant agreement No 757725).
\end{acknowledgments}

\end{document}